\documentclass[10pt]{article}

\date{March 2, 2017}
\def\eg{e.\kern -.4pt g., }
\def\Romannumeral#1{\uppercase\expandafter{\romannumeral #1}}

\usepackage{verbatim}
\usepackage{enumerate}

\usepackage{natbib}
\usepackage{lscape,rotating}
\usepackage{float,epsfig}
\usepackage{subfigure}
\usepackage{setspace}
\usepackage{multicol}

\usepackage{amsmath} 
\usepackage{amsfonts}
\usepackage{amssymb}

\usepackage{lscape,rotating}
\usepackage{float,epsfig}
\usepackage{subfigure}
\usepackage{setspace}
\usepackage{amsmath}
 \usepackage{amsmath,bm} 
\usepackage{amsfonts}
\usepackage{amssymb}

\newenvironment{proof}[1][Proof]{\begin{trivlist}
\item[\hskip \labelsep {\bfseries #1}]}{\end{trivlist}}

\begin{document}
\title{Sample size for comparing negative binomial rates  in noninferiority and equivalence trials with unequal follow-up times}
\author{Yongqiang Tang \\
yongqiang\_tang@yahoo.com\\
300 Shire Way, Lexington, MA 02421 \footnote{To appear in Journal of Biopharmaceutical Statistics}}
\maketitle

\begin{abstract}
We derive the sample size formulae
 for comparing  two negative binomial rates based on both the relative and absolute rate difference metrics  in noninferiority and equivalence trials with unequal follow-up times,
 and establish an approximate relationship between the sample sizes required for the treatment comparison based on the two treatment effect metrics. The proposed method allows
the dispersion parameter to vary by treatment groups.
The accuracy of these methods is assessed by simulations. 
It is demonstrated that ignoring the between-subject variation in the follow-up time by setting the follow-up time for all individuals to be the mean follow-up time may greatly underestimate the required size, resulting
in underpowered studies.  
Methods are provided for back-calculating the dispersion parameter based on the published summary results.
\end{abstract}


\setcounter{page}{0}

\setcounter{section}{0}
\setcounter{page}{1}
\section{Introduction}
Many clinical trials involve comparing the rate of events that may occur more than once in individual patients.
Examples include exacerbations in chronic obstructive pulmonary disease,
relapses in multiple sclerosis, tumor recurrence in bladder cancer,  seizures in epileptics, and hospitalizations.
When the event counts are analyzed using Poisson regression,   the observed variance is often larger than expected \citep{glynn:1996, wang:2009}, and
this phenomenon is called overdispersion. The  quasi-Poisson approach (i.e. Poisson regression with overdispersion adjustment), 
which simply inflates the variance obtained from the Poisson regression by a constant factor, may not be able to control the type I error well when there is a large variation in the follow-up time \citep{tang:2015}. This will be further illustrated  by simulation, and a theoretical justification will be provided  in Appendix \ref{typei}.
Negative binomial (NB) regression has been widely used to analyze recurrent events in recent years because it provides a  convenient way to
account for the overdispersion exhibited in the recurrent event data.

 The  noninferiority (NI) trials, commonly used in the drug development,  show that a new treatment is not materially less efficacious than a standard control treatment,
or more precisely that the new product is not worse than the active control by a pre-specified small amount called  NI margin   \citep{emanon:2005,fda:2010}.
The NI trial design is chosen if it would be unethical to run a placebo controlled trial 
or because the new treatment may offer important advantages over the standard treatment in terms of convenience of administration, improved safety, reduced cost, or better compliance \citep{hung:2007, hahn:2012}.

In equivalence trials, the objective is to demonstrate that the test product is not clinically  different from a standard control treatment \citep{liao:2015, alten:2015}. Equivalence trials
are often used in the development of the biosimilar product, which is a biological product that is highly similar to the reference product notwithstanding minor differences in
clinically inactive components \citep{fda:2012a}.

Sample size formulae have been developed for NI  and equivalence trials comparing rates of recurrent events by  \cite{cook:2007} and \cite{zhu:2016}.
Both approaches use the approximation  by setting the follow-up time for all patients to be the mean follow-up time.
 \cite{tang:2015} demonstrated both theoretically and numerically that ignoring the between-subject variation in the follow-up time
 leads to underpowered studies in superiority trials. We will show that the same conclusion holds for NI  and equivalence trials. 
In addition, \cite{cook:2007} did not assume the margin is fixed in the power calculation, and the interpretation of the $P$-value or type I error can be quite different
in the sense that it requires the assumption that both the historical  and NI trials can be repeated infinitely many times \citep{hung:2003,hung:2007}.
The recent regulatory guidelines \citep{emanon:2005,fda:2010} recommend the fixed margin approach. 

In this paper, we extend   \cite{tang:2015} sample size calculation method for comparing two NB rates in superiority trials
with unequal follow-up times  to NI and equivalence  trials. 
We compare the treatments on basis of both the absolute and relative rate difference metrics, and establish 
an approximate relationship between the required sample sizes on the two metrics.  The absolute rate difference metric was not studied 
by \cite{tang:2015}. We derive the power and sample size formulae for NI trials in  Section $2$, and for equivalence trials in Section $3$.
The proposed methods are assessed via simulation and compared with   \cite{zhu:2016} approach for two types of clinical trial designs. In one design, the planned treatment duration
is the same for all subjects. In the second design, subjects are enrolled at different calendar times, but administratively censored at the same calendar time. 
In Appendix \ref{eted}, we derive analytic expression for calculating the mean and variance of the follow-up time in the two designs.
We also extend the methodology by allowing the dispersion parameters to differ by treatment groups. 
The sample size determination requires information about the dispersion parameter $\kappa$, which is rarely reported in the medical literature. 
In Appendix \ref{estkappa}, we describe methods for back-calculating  $\kappa$ from published summary results.

\section{Sample size for NI trials}
The NB distribution  is the probability distribution of the number of failures $Y$ before  $\kappa^{-1}$  successes  in a series of independent Bernoulli trials with the same probability  $p$ of success
\begin{equation}\label{nbdist}
 \Pr(Y=y) =
 \frac{\Gamma(y+1/\kappa)}{y!\,\Gamma(1/\kappa)}  p^{1/\kappa} (1- p)^{y}.
\end{equation}
For the analysis of overdispersed count data, it is convenient to use the parametrization: $\mu=(1-p)/(\kappa p)$ is the mean, and $\kappa$ is the dispersion parameter. Note that $p=1/(\kappa \mu+1)$.
The NB distribution can be derived as a gamma mixture of Poisson distribution.
If $Y$ is Poisson distributed with mean $\varepsilon \mu$, and $\varepsilon$ is gamma distributed with mean $1$ and variance $\kappa$, the
marginal distribution of $Y$ is a NB distribution, and we will denote it by $Y\sim \mathcal{NB}(\mu,\kappa)$. 
 This representation does not require $\kappa^{-1}$  to be an integer. 
The random effect   $\varepsilon$ can also be modeled by other distributions such as the log-normal or inverse Gaussian distributions, but the gamma mixing distribution
is the most commonly used because the resulting marginal distribution has a closed-form expression \citep{lawless:1987}.
The random effect $\varepsilon$ captures the between-subject heterogeneity in event rates,
and   $\kappa$ measures the degree of heterogeneity. Including important risk factors in the model may
reduce heterogeneity. 
The NB distribution tends to fit the overdispersed count data better than the Poisson distribution \citep{glynn:1996, wang:2009}, and its mean $\mu$  is always less than its variance 
$\mu+\kappa\mu^2$ \citep{tang:2015}.

Suppose in a trial, $n$ subjects are randomized to receive an active ($g=1$) or control ($g=0$)
treatment. Let $t_{gj}$ denote the follow-up time and $y_{gj}$ the number of observed events
for subject $j=1,\ldots,n_g$ in treatment group $g$. We assume that the event rate $\lambda_g=\exp( \gamma_g)$ in each treatment group
is constant over time, and $y_{gj} \sim \mathcal{NB}(\lambda_g t_{gj},\kappa) $. The log-likelihood function can be written as
\begin{equation}\label{logl}
\ell= \sum_{g=0}^1\sum_{j=1}^{n_g}\left[\log\frac{\Gamma(y_{gj}+1/\kappa)}{\Gamma(1/\kappa)} + y_{gj} \log(\kappa \lambda_g t_{gj}) -(y_{gj}+\kappa^{-1}) \log(1+\kappa\lambda_gt_{gj})\right].
\end{equation}
Based on the analytic result of \cite{lawless:1987}, \cite{tang:2015} showed that the maximum likelihood estimates (MLE) $\hat\gamma_0$, $\hat\gamma_1$ and $\hat\kappa$
are asymptotically independent, and the variance of the log-relative risk estimate $\hat\beta=\log(\hat\lambda_1/\hat\lambda_0)=\hat\gamma_1-\hat\gamma_0$ derived from the expected Fisher information matrix is given 
by 
\begin{equation}\label{varbeta}
\text{var}(\hat\gamma_g)=\left(\sum_{j=1}^{n_g} \frac{ \lambda_g t_{gj}}{1+\kappa \lambda_g t_{gj}}\right)^{-1}\approx \frac{1}{np_gd_g} \text{ and }
\text{var}(\hat\beta)=\sum_{g=0}^1\text{var}(\hat\gamma_g) \approx \frac{1}{ n}\left [\frac{1}{d_0p_0}+\frac{1}{d_1p_1}\right],
\end{equation}
where $d_g=\text{E}[\lambda_g t_{gj}/(1+\kappa \lambda_g t_{gj})]$, and $p_g=n_g/n$ is the proportion of subjects randomized to treatment group $g$.
The variance of $\hat\beta$ can also be obtained from the observed information matrix. The two variance estimates are asymptotically equivalent. The $100(1-\alpha)\%$ confidence interval (CI)  is 
$$[c_l,c_u]= [\hat\beta -z_{1-\alpha/2}\sqrt{\widehat{\text{var}}(\hat\beta)}, \hat\beta +z_{1-\alpha/2}\sqrt{\widehat{\text{var}}(\hat\beta)}]$$
 for $\beta$, and $[\exp(c_l),\exp(c_u)]$ for the event rate ratio $\lambda_1/\lambda_0$, where $\widehat{\text{var}}(\cdot)$ is the variance estimated at the MLE,
and $z_p$ is the $p$th percentile of the standard normal distribution $N(0,1)$.

\subsection{Sample size for the rate ratio effect measure}
Suppose a lower event rate is desirable. 
In the NI trial, the objective is to demonstrate that the experimental treatment is no worse than the active comparator by $M_{r0}$, where $M_{r0}>1$ is the pre-specified NI margin on the rate ratio.
The hypothesis can be expressed as
$$ H_0: \frac{\lambda_1}{\lambda_0}\geq M_{r0}  \text{ {\it vs }} H_1: \frac{\lambda_1}{\lambda_0} < M_{r0}.$$
The margin $M_{r0}$ is generally chosen to be close to $1$ in order to demonstrate that the new treatment is not materially inferior to the active comparator. 
Please refer to \cite{ fda:2010}, \cite{emanon:2005} and \cite{hung:2003} for more details on the specification and interpretation of the NI margin.
In the fixed margin approach, we compare the upper limit of the CI for $\lambda_1/\lambda_0$ with  $M_{r0}$. 
 The noninferiority of the experimental treatment to the active comparator can be claimed if $\exp(c_u)<M_{r0}$ or equivalently if $c_u=\hat\beta +z_{1-\alpha/2}\sqrt{\widehat{\text{var}}(\hat\beta)}<\log(M_{r0})$. 
The power of the test is 
\begin{equation*}
 P=\Pr( \hat\beta +z_{1-\alpha/2}\sqrt{\widehat{\text{var}}(\hat\beta)} < \log(M_{r0})) =\Pr\left( Z < \frac{-z_{1-\alpha/2}\sqrt{\widehat{ \text{var}}(\hat\beta) }-\beta+\log(M_{r0}) }{\sqrt{\text{var}(\hat\beta)}} \right),
\end{equation*}
which can be expressed as
\begin{equation}\label{pow1}
 P\approx \Phi\left(\frac{\sqrt{n}|\beta^*|}{\sqrt{ (d_0p_0)^{-1}+(d_1p_1)^{-1}}}-z_{1-\alpha/2}\right) = \Phi\left(\frac{\sqrt{n}|\log(M_{r0})-\log(\frac{\lambda_1}{\lambda_0})|}{\sqrt{ (d_0p_0)^{-1}+(d_1p_1)^{-1}}}-z_{1-\alpha/2}\right),
\end{equation}
where $\beta^*=\log(M_{r0}\lambda_0/\lambda_1)$ and $Z=(\hat\beta-\beta)/\sqrt{\text{var}(\hat\beta)}$ is asymptotically distributed as  $N(0,1)$. Equation \eqref{pow1} is also valid if 
$M_{r0}<1$, and   the objective is to demonstrate that the event rate in the experimental arm is not materially lower than  the control rate.
Inverting \eqref{pow1} yields the total  sample size
\begin{equation}\label{sizesnon}
n_{r} =\left[\frac{1}{d_0p_0}+\frac{1}{d_1p_1}\right] f,
\end{equation}
where $f= (z_{1-\alpha/2}+z_{P})^2/\beta^{*^2}$. 
The formula of  \cite{friede:2010} is a special case of \eqref{sizesnon}  when all subjects  have the same follow-up time.
When $M_{r0}=1$, \eqref{sizesnon} reduces to the sample size formula of \cite{tang:2015} for superiority trials. 
The superiority and NI trials differ in  the assumption on the relative efficacy of the two treatments. 
In superiority trials, $\lambda_1/\lambda_0$ is expected to be far below $1$. In NI trials, 
the active and control treatments are expected to have similar effects (i.e. $\lambda_1/\lambda_0\approx 1$).

The sample size  for NI trials can be determined by adapting the formulae for superiority trials  \citep{friede:2010} with $\beta^{*2}=\log^2(M_{r0}\lambda_0/\lambda_1)$
 replacing $\beta^2=\log^2(\lambda_0/\lambda_1)$. In this approach, $d_0$, $d_1$ and the variance of the treatment effect are still calculated based on the true event rates.
One shall be cautious in using this adapted approach by avoiding the possibility of modifying the true rate as $\lambda_1=\exp(\beta^*)\lambda_0$.  
Such a mistake was made in \cite{zhu:2016}, leading him to express concerns about the inaccuracy of this adapted approach. 
Let take the case reported on row $8$, table 5 in the supplementary material of \cite{zhu:2016} as an example.
Suppose $\kappa=0.5$, $\lambda_0=\lambda_1=1$,  $t_{gj}\equiv 1$ (no dropout), $p_0=p_1=1/2$, and $M_{r0}=1.3$. 
By \eqref{sizesnon},  $686$ subjects ($343$ per arm) are needed to achieve a power of $P=80\%$ at $\alpha=0.05$. 
The  power estimated from $40,000$ simulated trials is $80.43\%$. 
 \cite{zhu:2016} got a size estimate of $314$ subjects per group  using 
method  $3$ of  \cite{zhu:2014} for superiority trials by replacing $\exp(\beta)$ by $M_{r0}$ (i.e. $\beta^*=\log(M_{r0})$ since $\lambda_0=\lambda_1$).
 \cite{zhu:2016}  estimate was  wrong because the true  rates were  implicitly modified  as $(\lambda_0,\lambda_1)=(1,1.3)$ in his calculation.
The resulting size is also   $343$ subjects per arm if \cite{zhu:2014} formula for superiority trials is appropriately applied. 
The simulation studies in both \cite{friede:2010} and this paper demonstrate the accuracy of \eqref{sizesnon} or equivalently the adapted method in moderate to large samples.

 \cite{tang:2015} derived the lower  and upper  bounds for $d_g$
\begin{equation}\label{boundd}
d_{g_l} = \frac{\lambda_g \nu_{t_g}^2}{\nu_{t_g}+\kappa  \lambda_g \text{E}(t_{gj}^2)}\leq d_g \leq d_{g_u} =  \frac{\lambda_g \nu_{t_g}}{1+\kappa \lambda_g\nu_{t_g}},
\end{equation}
where $\nu_{t_g}=\text{E}(t_{gj})$ is the mean follow-up time in group $g$.
Replacing $d_g$ by $d_{g_u}$ and $d_{g_l}$ in \eqref{sizesnon} yields respectively the lower and upper bounds for $n_{r}$.
Similar bounds on the power can be obtained  by replacing $d_g$ by $d_{g_l}$ and $d_{g_u}$ in \eqref{pow1}.

The lower size bound is the required size when all subjects in treatment group $g$ are followed for the
same time $t_{gj}=\nu_{t_g}$, and  
 can be decomposed into two terms
\begin{equation}\label{low}
n_{rl} =  \left[ \frac{1}{p_0\lambda_0 \nu_{t_0}} + \frac{1}{p_1\lambda_1 \nu_{t_1}} \right]\,f + \left(\frac{1}{p_0}+\frac{1}{p_1}\right)\kappa \,f.
\end{equation} 
The first term is the required size ignoring overdispersion (i.e. the count data follow Poisson distribution), and the second term 
corrects for overdispersion. In the upper size bound, another  term is added to account for variation in the duration of the follow-up
\begin{equation}\label{upper1}
n_{ru}  = n_{rl}+\kappa f \left[ \frac{1}{p_0}\text{CV}_{0}^2 +\frac{1}{p_1}\text{CV}_{1}^2\right],
\end{equation} 
where $\text{CV}_{g}=\sqrt{\text{var}(t_{gj})}/{\nu_{t_{gj}}}$ is the coefficient of variation for $t_{gj}$. Note that $n_{ru}$ can be  bounded by
$$n_{ru}\leq \tilde{n}_{ru} = n_{rl}+ \kappa f [\frac{t_{m_0}-\nu_{t_0}}{\nu_{t_0}}+\frac{t_{m_1}-\nu_{t_1}}{\nu_{t_1}}],$$
where $t_{m_g}$ is the maximum follow-up time in treatment group $g$.

The power and sample size calculation requires $d_g$, $\nu_{t_g}=\text{E}(t_{gj})$, $\text{E}(t_{gj}^2)$, and the corresponding analytic formulae
are given in Appendix \ref{eted} for two types of designs considered in Section $2.3.1$. 
The estimate of the dispersion parameter $\kappa$ is rarely reported in the medical literature. Appendix \ref{estkappa} illustrates how to use \eqref{boundd} to back-calculate $\kappa$ from 
the summary results on the event rate and rate ratio.

\subsection{Sample size for the absolute rate difference measure}
Now suppose we want to compare the treatments based on the  absolute rate difference measure. The objective is 
 to show that the active rate is no worse than the control rate by $M_{d0}$, where $M_{d0}$ is the NI margin on the absolute rate difference  metric. The hypothesis can be expressed as
$$ H_0: \lambda_d= \lambda_1- \lambda_0\geq M_{d0}  \text{ {\it vs }} H_1:  \lambda_d< M_{d0}.$$
The variance of $\hat\lambda_d=\hat\lambda_1-\hat\lambda_0=\exp(\hat\gamma_1)-\exp(\hat\gamma_0)$ can be derived from  \eqref{varbeta} via the delta method
$$\text{var}(\hat\lambda_g)=  \lambda_g^2\text{var}(\hat\gamma_g)  \text{ and }
\text{var}(\hat\lambda_d)=\sum_{g=0}^1\text{var}(\hat\lambda_g) =\frac{1}{n}\left[\frac{\lambda_0^2}{d_0p_0}+ \frac{\lambda_1^2}{d_1p_1}\right].$$ 
The $100(1-\alpha)\%$ CI for $\hat\lambda_d$ is 
$$[c_{l_d},c_{u_d}]=[\hat\lambda_d-z_{1-\alpha/2} \sqrt{\widehat{\text{var}}(\hat\lambda_d)}, \hat\lambda_d+z_{1-\alpha/2} \sqrt{\widehat{\text{var}}(\hat\lambda_d)}].$$
The noninferiority can  be claimed if $c_{u_d}<M_{d0}$.  The power of the test is 
\begin{equation*}
 P=\Pr(c_{u_d}<M_{d0})
=\Pr\left( \frac{\hat\lambda_d-\lambda_d}{\sqrt{ \text{var}(\hat\lambda_d)}}< \frac{M_{d0}-\lambda_d-z_{1-\alpha/2}\sqrt{\widehat{ \text{var}}(\hat\lambda_d) }}{\sqrt{ \text{var}(\hat\lambda_d)}} \right),
\end{equation*}
which can be written as 
\begin{equation}\label{pow2}
 P = \Phi\left(\frac{\sqrt{n}|M_{d0}+\lambda_0-\lambda_1|}{\sqrt{\frac{\lambda_0^2}{d_0p_0}+ \frac{\lambda_1^2}{d_1p_1}}}-z_{1-\alpha/2}\right).
\end{equation}
Equation \eqref{pow2} is also valid if 
$M_{d0}<0$ and   the objective is to demonstrate that the event rate in the experimental arm is not materially lower than that in  the control arm.
Inverting \eqref{pow2} yields the sample size
\begin{equation}\label{sizenondiff}
 n_d=\left(\frac{\lambda_0^2}{p_0d_0}+\frac{\lambda_1^2}{p_1d_1}\right)  \frac{(z_{1-\alpha/2}+z_P)^2}{(M_{d0}+\lambda_0-\lambda_1)^2}.
\end{equation}
When $M_{d0}=0$, \eqref{sizenondiff} reduces to the sample size formula for testing the absolute rate difference in superiority trials. 
Replacing $d_g$ by $d_{g_u}$ and $d_{g_l}$ in \eqref{sizenondiff} yields respectively the lower $n_{dl}$ and upper $n_{du}$ bounds for $n_{d}$.

In Appendix \ref{proof}, we show that \eqref{sizesnon} and \eqref{sizenondiff} generally produce similar sample size estimate if the effects of the two treatments do not differ too much,
 and
\begin{equation}\label{equiv}
M_{d0}= \bar\lambda \log(M_{r0}).
\end{equation}
where $\bar\lambda=\sqrt{\lambda_0\lambda_1}=\lambda_0\exp(\beta/2)$ is the geometric mean of $\lambda_0$ and $\lambda_1$.

\subsection{Numerical Examples}

\subsubsection{Assessment of type I error rate}
We perform simulations to assess the type I error of the Wald-CI-based approach in the NB regression and quasi-Poisson regression with overdispersion adjustment.
Two types of designs are considered. 
In design $1$, the planned treatment duration is $\tau_c$ years for all patients, and the loss to follow-up is assumed to be exponentially distributed with mean $\delta^{-1}$ years, and 
independent of the recurrent event process.  In design 2, we assume patients are enrolled during an accrual period of $\tau_a$ years, and 
followed for an additional $\tau_c$ years after the closure of recruitment \citep{cook:1995}. 
The patient entry time $e_{gj}$ is distributed with density function
\begin{equation}\label{entry}
g(e_{gj}) = \frac{\eta \exp(-\eta e_{gj})}{1-\exp(-\eta\tau_a)},\quad 0 \leq e_{gj}\leq \tau_a,
\end{equation}
given by \cite{lachin:1986}.
 The entry distribution is convex (faster patient entry at the beginning) if $\eta>0$, and
concave (lagging patient entry) if $\eta<0$, and uniform $g(e_{gj})=1/\tau_a$ if $\eta\rightarrow 0$. We assume uniform entry in the simulation.
The  loss to follow-up distribution is the same as design $1$. 
Since all subjects will be administratively censored at time $\tau=\tau_a+\tau_c$, but enter the study at different time, there is greater between-subject variation in the follow-up time in design $2$ than in design $1$.
Appendix \ref{eted} provides  analytic formulae for calculating $d_g$, $\nu_{t_g}=\text{E}(t_{gj})$ and $\text{E}(t_{gj}^2)$ for the two designs.

In design $1$, we assume $\tau_c=2$ years, and   the overall dropout rate at year 2 is $25\%$ ($\delta=0.1438$). 
In design $2$, we fix $\tau_a=\tau_c=2$ years, and the loss to follow-up is exponentially distributed with mean $1/\delta=5$ years.
In both designs, we set $(\lambda_0,\kappa)=(0.6,1)$ or $(0.9,1.5)$, and $M_{r0}=1.2$ or $1.3$. Equal treatment allocation ($p_0=p_1=1/2$) is assumed in the simulation.
 For the test based on the rate ratio effect metric, 
 we assume $\lambda_1=\lambda_0M_{r0}$ in  simulating the data, and 
the total sample size is estimated using \eqref{sizesnon}  at the target $P=80\%$ power and a two-sided $\alpha=0.05$ significance level under the assumption $\exp(\beta)=\lambda_1/\lambda_0=0.65, 0.8, 0.9, 1, 1.05$. 
For the test of the absolute rate difference, we set $\lambda_1=\lambda_0+M_{d0}$ in simulating the data, 
and the sample size is determined by  \eqref{sizenondiff} at $P=80\%$ and $\alpha=0.05$ under the assumption $\exp(\beta)=\lambda_1/\lambda_0=0.65, 0.8, 0.9, 1, 1.05$, 
where $M_{d0}=\lambda_0\exp(\beta/2) \log(M_{r0})$.  We do not consider the case when  $\exp(\beta)>1.05$ partially because the resulting sample size could too large to be of practical interest, and partially because
 the performance of the NB regression is expected to become better as the sample size increases.
In all cases, $10,000$ trials are simulated, and 
there is $>95\%$ chance that the empirical one-sided type I error estimate (its standard error  is about $\sqrt{0.025*0.975/10000} \approx 0.16\%$) is within $0.32\%$ of the true error rate.

The results are reported in table \ref{type1_sim1} and \ref{type1_sim2}  respectively for design $1$  and $2$.  The test based on the rate ratio metric generally provides
 a better  type I error control than the test of the absolute rate difference. 
For the NB regression, the empirical estimate of the one-sided type I error  is generally close to the nominal $0.025$ level.
In the worst case, the empirical  type I error estimate reaches $2.81\%$ for the test based on the rate ratio, and 
$2.97\%$  for the test based on the absolute rate difference.

The quasi-Poisson regression  provides quite poor control of the type I error in NI trials, and similar phenomenon was observed in superiority trials \citep{tang:2015}.
In NI trials, the empirical one-sided type I error estimate is
in the $3.0-4.0\%$ range for all cases in design $2$ because of the large variation in the follow-up time, and above $2.80\%$
 in most cases for design $1$. In Appendix \ref{typei}, we show that the variance of the treatment effect estimate is underestimated under $H_0$ in the quasi-Poisson regression if there is a large variation in $t_{gj}$'s, and this explains why the type I error is inflated in this approach. 

\subsubsection{Assessment of  power and sample size}
The  set up for assessing the performance of the power and sample size formulae is
 similar to that for the type I error assessment except that we assume $\lambda_1=\lambda_0\exp(\beta)$ in simulating data for both
tests based on the relative and absolute rate difference, and the two margins satisfy $M_{d0}=\sqrt{\lambda_0\lambda_1}\log(M_{r0})$.
Since the quasi-Poisson regression
could not control the type I error well, its performance on power will not be assessed. We will compare \eqref{sizesnon} with \cite{zhu:2016} method. In the latter approach,  the follow-up time for all individuals is set to be the mean follow-up time, 
\begin{equation}\label{zhunon}
n_{zr} =  \frac{\left[ z_{1-\alpha/2}\sqrt{\tilde{V}_0} + z_P\sqrt{V_1} \right]^2}{\log^2(M_{r0}\lambda_0/\lambda_1)},
\end{equation} 
where $\nu_t=\nu_{t_0}=\nu_{t_1}$, 
$V_1= (p_0^{-1}+p_1^{-1})\kappa + ({p_0\lambda_0\nu_{t}})^{-1} +(p_1\lambda_1\nu_{t})^{-1}$,
$\tilde{V}_0= (p_0^{-1}+p_1^{-1})\kappa + ({p_0\tilde\lambda_0\nu_{t}})^{-1} +(p_1\tilde\lambda_1\nu_{t})^{-1}$,
$\theta=p_1/p_0$, 
$a=-\kappa\nu_t M_{r0} (1+\theta)$, $b=\kappa \nu_t(\lambda_0 M_{r0}+\theta\lambda_1)-(1+\theta M_{r0})$, 
$c=\lambda_0+\theta\lambda_1$, and 
$\tilde\lambda_0 =(-b-\sqrt{b^2-4ac})/(2a)$ and $\tilde\lambda_1=M_{r0}\tilde\lambda_0$ are the approximate MLE under $H_0: \lambda_1=M_{r0}\lambda_0$.

The results are reported in table \ref{power_sim1} and \ref{power_sim2}  respectively for design $1$  and $2$. 
As expected, the estimated size $n_{d}$ for the test of the rate difference is generally close to the size $n_r$ for the test based on the rate ratio especially when 
$\exp(\beta)=\lambda_1/\lambda_0$ is near $1$.  If the experimental treatment is much more effective than the control treatment, $n_{d}$ tends to be slightly larger than $n_r$.
For the test based on the rate ratio effect metric, the simulated power (SIM) at $n_{r}$ is within $1\%$ of the nominal power in nearly all cases for both designs.

We evaluate the power of  the test of the absolute rate difference at both $n_d$ and $n_r$. 
The SIM at $n_{d}$ is generally close to the targeted $80\%$ level, and the power approximation slightly deteriorates when the experimental treatment is much more effective than the active control 
(i.e. $\exp(\beta)=\lambda_1/\lambda_0=0.65$ and $0.8$), and the SIM   
deviates from the nominal level by $2.48\%$ in the worst case.  The SIM at $n_r$ is closer to the targeted level than the SIM at $n_d$ when $\exp(\beta)=\lambda_1/\lambda_0=0.65$ and $0.8$. 
This suggests that  the sample size for the test of the absolute rate difference can be calculated based on either  \eqref{sizenondiff} or \eqref{sizesnon} by setting $M_{r0}=\exp(M_{d0}/\bar\lambda)$, and 
the latter approach may be preferred.

 \cite{zhu:2016} estimate $n_{zr}$ given in \eqref{zhunon} is generally close to the lower size bound $n_{rl}$, and $n_{zr} < n_{rl}$ when the experimental treatment is much more effective than the active control 
($\exp(\beta)=\lambda_1/\lambda_0$ is far below $1$).
The use of  the lower size bound $n_{rl}$ or Zhu's estimate $n_{zr}$ by ignoring the variability in the duration of the follow-up
underestimates the required size, and the amount of underestimation ranges from $3-6\%$ for design $1$, and $7-10\%$ for design $2$. 
The upper size bounds overestimate the sample size by $1-2.5\%$ in design $1$, and by $3-9\%$ in design $2$.

\subsection{Extension to dispersion parameter heterogeneity}
So far we assume a common dispersion parameter $\kappa$  across treatment groups. In practice, the dispersion parameter could differ by treatments. If the assumption of a common dispersion parameter
is relaxed in the analysis, the corresponding power and sample size formulae remain almost unchanged except that one needs to replace
$\kappa$ in $d_g$, $d_{g_u}$ and $d_{g_l}$ by the treatment specific dispersion parameter $\kappa_g$. 
This modification is also suitable for superiority trials \citep{tang:2015} since superiority trials can be viewed as the special cases of NI trials
 when $M_{r0}=1$ and $M_{d0}=0$.

We perform a small simulation study to assess the accuracy of the  power and sample size formulae under heterogeneous dispersion. Only design $1$ is considered, where $\tau_c=2$, $p_0=p_1=1/2$, 
and the overall dropout rate is $25\%$ ($\delta=0.1438$).
We set $\lambda_0=0.6$ or $1$, $\exp(\beta)=\lambda_1/\lambda_0=0.8$, $0.9$ or $1$, $(\kappa_0,\kappa_1)=(2,1)$, $(1,2)$, $(2,0.5)$ or $(0.5,2)$, 
 $M_{r0}=1.2$ or $1.3$, and $M_{d0}=\sqrt{\lambda_0\lambda_1} \log(M_{r0})$. 

In the analysis of simulated data, we fit a separate NB regression for each treatment group, and calculate the test statistic based on the analytic formula. 
 This  analysis strategy works if the model does not include other covariates except the treatment status.
The results at $M_{r0}=1.3$ are displayed in Table \ref{power_herdisp}. The performance at $M_{r0}=1.2$ is similar, and hence omitted. 
For the test based on the event rate ratio,  the SIM at $n_r$ is generally close to the $80\%$ nominal level.  The power approximation for the test based on the absolute difference 
may slightly deteriorate, and the SIM at $n_d$ can be off the targeted power by about $4\%$. Again, the results indicate that it might be better to use \eqref{sizesnon} to calculate 
the sample size for the test of the absolute rate difference by setting $M_{r0}=\exp(M_{d0}/\bar\lambda)$.

\section{Sample size for equivalence trials}

\subsection{Sample size for the rate ratio effect measure}
The purpose of an equivalence trial is to demonstrate that the test product is neither superior nor inferior to the reference product.
The hypothesis can be written as
$$ H_0: \frac{\lambda_1}{\lambda_0}\leq M_{rl} \text{ or }  \frac{\lambda_1}{\lambda_0}\geq M_{ru} \text{ {\it vs }} H_1: M_{rl}< \frac{\lambda_1}{\lambda_0} < M_{ru},$$
where $M_{rl}<1$  and $M_{ru}>1$ are the pre-specified lower and upper equivalence boundaries on the rate ratio.

The two treatments are not clinically different  if the whole CI  of  $\exp(\beta)=\lambda_1/\lambda_0$ lies completely within $[M_{rl}, M_{ru}]$, or equivalently if the whole CI 
of $\beta$ falls completely within  $[\log(M_{rl}), \log(M_{ru})]$. The power is 
\begin{eqnarray}\label{powerequiv}
\begin{aligned}
P &=  \Pr( \hat\beta+z_{1-\alpha/2} \sqrt{\widehat{ \text{var}}(\hat\beta)}<\log(M_{ru}) \text{ and }   \hat\beta-z_{1-\alpha/2} \sqrt{\widehat{ \text{var}}(\hat\beta)}>\log(M_{rl}) ) \\
&= \Pr\left(    \frac{z_{1-\alpha/2}\sqrt{\widehat{ \text{var}}(\hat\beta) }-\beta+\log(M_{rl}) }{\sqrt{\text{var}(\hat\beta)}} <     Z < \frac{-z_{1-\alpha/2}\sqrt{\widehat{ \text{var}}(\hat\beta) }-\beta+\log(M_{ru}) }{\sqrt{\text{var}(\hat\beta)}} \right)\\
& \approx \max\left\{\Phi\left(\frac{\sqrt{n}\log(M_{ru}\lambda_0/\lambda_1)}{\sqrt{ (d_0p_0)^{-1}+(d_1p_1)^{-1}}}-z_{1-\alpha/2}\right) -
         \Phi\left(\frac{\sqrt{n}\log(M_{rl}\lambda_0/\lambda_1)}{\sqrt{ (d_0p_0)^{-1}+(d_1p_1)^{-1}}}+z_{1-\alpha/2}\right), 0\right\}.
\end{aligned}
\end{eqnarray}
There is no closed-form sample size formula except in few special cases.  When $\log(M_{ru}\lambda_0/\lambda_1) = -\log(M_{rl}\lambda_0/\lambda_1)$, the  sample size is given by
\begin{equation}\label{sizeequiv}
n= \left [\frac{1}{d_0p_0}+\frac{1}{d_1p_1}\right]\,\frac{(z_{1-\alpha/2}+z_{(1+P)/2})^2}{\log^2(M_{ru}\lambda_0/\lambda_1)} = \left [\frac{1}{d_0p_0}+\frac{1}{d_1p_1}\right]\,\frac{(z_{\alpha/2}+z_{(1-P)/2})^2}{\log^2(M_{ru}\lambda_0/\lambda_1)}.
\end{equation}
Let $\Delta_{min}=\min[\log(M_{ru}\lambda_0/\lambda_1), -\log(M_{rl}\lambda_0/\lambda_1)]$ and 
$\Delta_{max}=\max[\log(M_{ru}\lambda_0/\lambda_1), -\log(M_{rl}\lambda_0/\lambda_1)]$.
In Appendix \ref{proof}, we show   that    the sample size is bounded by
\begin{equation}\label{sizeequiv2}
 \left [\frac{1}{d_0p_0}+\frac{1}{d_1p_1}\right]\,\frac{(z_{1-\alpha/2}+z_{(1+P)/2})^2}{\Delta_{max}^2} \leq n \leq   
\left[\frac{1}{d_0p_0}+\frac{1}{d_1p_1}\right]\,\frac{(z_{1-\alpha/2}+z_{(1+P)/2})^2}{\Delta_{min}^2}. 
\end{equation}
The sample size can be approximated by the upper bound in \eqref{sizeequiv2}  if $\Delta_{min}$ is sufficiently close to $\Delta_{max}$. 

When $M_{rl} \lambda_0/\lambda_1$ is sufficiently  smaller than $1$,  \eqref{powerequiv} can be approximated by \citep{chow:2001, chow:2008}
$$ P  \approx  \Phi\left(\frac{\sqrt{n}\log(M_{ru}\lambda_0/\lambda_1)}{\sqrt{ (d_0p_0)^{-1}+(d_1p_1)^{-1}}}-z_{1-\alpha/2}\right) $$
and the sample size can be approximated by the size for a NI trial with margin $M_{r0}=M_{ru}$
$$ n\approx \left[\frac{1}{d_0p_0}+\frac{1}{d_1p_1}\right]\,\frac{(z_{1-\alpha/2}+z_{P})^2}{\log^2(M_{ru} \lambda_0/\lambda_1)}.$$ 
Similar approximation may be obtained when $M_{ru}\lambda_0/\lambda_1$ is sufficiently larger than $1$.

In general,  sample size may be obtained by inverting \eqref{powerequiv}  numerically  (e.g. bisection method), which is the smallest integer at which
 the  power in \eqref{powerequiv} is not less than the target power.  

\subsection{Sample size for the absolute rate difference measure}
On the absolute rate difference metric, the hypothesis is
$$ H_0: \lambda_d=\lambda_1-\lambda_0\leq M_{dl} \text{ or } \lambda_d \geq M_{du} \text{ {\it vs }} H_1:  M_{dl}< \lambda_d< M_{du},$$
where $M_{dl}<0$ and $M_{du}>0$ are the pre-specified lower and upper equivalence boundaries on the absolute rate difference metric. We can derive the power 
\begin{eqnarray}\label{powerequivd}
\begin{aligned}
P &=  \Pr( c_{l_d}> M_{dl} \text{ and } c_{u_d} <M_{du})\\
&=\Pr\left( \frac{M_{dl}-\lambda_d+z_{1-\alpha/2}\sqrt{\widehat{ \text{var}}(\hat\lambda_d) }}{\sqrt{ \text{var}(\hat\lambda_d)}}< \frac{\hat\lambda_d-\lambda_d}{\sqrt{ \text{var}(\hat\lambda_d)}}< \frac{M_{du}-\lambda_d-z_{1-\alpha/2}\sqrt{\widehat{ \text{var}}(\hat\lambda_d) }}{\sqrt{ \text{var}(\hat\lambda_d)}} \right)\\
& \approx \max\left\{\Phi\left(\frac{\sqrt{n}(M_{du}+\lambda_0-\lambda_1)}{\sqrt{\frac{\lambda_0^2}{d_0p_0}+ \frac{\lambda_1^2}{d_1p_1}}}-z_{1-\alpha/2}\right) -\Phi\left(\frac{\sqrt{n}(M_{dl}+\lambda_0-\lambda_1)}{\sqrt{\frac{\lambda_0^2}{d_0p_0}+ \frac{\lambda_1^2}{d_1p_1}}}+z_{1-\alpha/2}\right),0\right\}.
\end{aligned}
\end{eqnarray}
Sample size can be calculated by numerically inverting \eqref{powerequivd}. In the special case when  $M_{du}+\lambda_0-\lambda_1=-(M_{dl}+\lambda_0-\lambda_1)$, we have
$$ n= \left[\frac{\lambda_0^2}{p_0d_0}+\frac{\lambda_1^2}{p_1d_1}\right]  \frac{(z_{1-\alpha/2}+z_{(1+P)/2})^2}{(M_{du}+\lambda_0-\lambda_1)^2}.$$

When the effects of the two treatments do not differ too much, and the margins satisfy  $M_{dl}=\bar\lambda\log(M_{rl})$ and $M_{du}=\bar\lambda\log(M_{ru})$,
the power estimates based on rate ratio metric (i.e. via \eqref{powerequiv}) and the absolute rate difference  metric (i.e. via \eqref{powerequivd}) are close, and therefore the corresponding sample
size estimates are close. The proof is similar to that for \eqref{equiv}.

As in Section $2$, replacing $d_g$ by $d_{g_l}$ or $d_{g_u}$ in the power  formulae for tests based on both the absolute and relative rate difference metrics yields
the  lower and upper power bounds. The sample size bounds can be obtained by numerically inverting the power bounds.
If the dispersion parameter differs by treatment, the corresponding power and sample size formulae remain almost unchanged except that one needs to replace
$\kappa$ in $d_g$, $d_{g_u}$ and $d_{g_l}$ by the treatment specific dispersion parameter $\kappa_g$. 

\subsection{Numerical examples}
We perform a small simulation study to assess the accuracy of the  power and sample size formulae for equivalence trials. 
 We fix $\tau_c=2$ years in design $1$, and $\tau_a=\tau_c=2$ years in design $2$. The distribution of the loss to follow-up time  is the same
as that in simulation $1$ in Section $2.3.1$. In both designs, we set $(\lambda_0,\kappa)=(0.6,1)$ or $(0.9,1.5)$ and $\exp(\beta)=\lambda_1/\lambda_0=1$ or $1.05$.
We choose the margin as $M_{ru}=1/M_{rl}=1.3$ on the rate ratio metric, and $M_{du}=-M_{dl}=\sqrt{\lambda_0\lambda_1}\log(M_{ru})=\lambda_0\exp(\beta/2)\log(M_{ru})$ on the absolute rate difference metric.

The simulated powers are within $1\%$ of the nominal level for tests based on both absolute and relative rate difference metrics in all cases. 
The use of  the lower size bound $n_{rl}$ or Zhu's estimate $n_{zr}$ by ignoring the variability in the duration of the follow-up underestimates the required size, and 
the degree of underestimation is similar to that for NI trials reported in Section $2.3.2$. 

The  equivalence trial often requires larger sample size than a NI trial  \citep{liao:2015}. Below is an example.
At $(\lambda_0,\lambda_1/\lambda_0,\kappa)=(0.6,1,1)$ in design $1$, it needs $928$ subjects (reported in Table \ref{power_sim1})
 to demonstrate with $80\%$ power that the test treatment is not inferior to the reference product at a margin of $M_{r0}=1.3$ on the rate ratio metric, while
$1242$ subjects are required to show that the test product is neither inferior nor superior to the reference product at the margin of $M_{ru}=1/M_{rl}=1.3$.
The equivalence trial requires $1242/928-1=33.8\%$ more subjects than the NI trial. 
The difference in sample sizes between the NI and equivalence trials becomes smaller ($1384$ vs $1435$ subjects) at  $(\lambda_0,\lambda_1/\lambda_0,\kappa)=(0.6,1.05,1)$, and 
becomes larger ($668$ vs $1469$ subjects) at  $(\lambda_0,\lambda_1/\lambda_0,\kappa)=(0.6,0.952,1)$. 

We also conduct a simulation study to assess the type I error of the equivalence tests. The empirical type I error is generally close to the  nominal  level. 
The result is not reported due to limited space.

\section{Discussion}
We derive the power and sample size formulae for comparing two NB rates based on the absolute and relative rate difference  in NI and equivalence trials.
The sample size formula for superiority trials can be viewed as a special case of the formula for NI trials by setting $M_{r0}=1$ and $M_{d0}=0$. 
 The assumption of a common dispersion parameter
across treatments can be relaxed. 
The accuracy of the proposed methods is demonstrated by simulations
 in moderate to large samples. We show  that the sample size  in NI and equivalence trials will be underestimated if one 
 ignores the variation in the follow-up time by setting the follow-up time for all individuals to be the mean follow-up time
and estimates the sample size using either the lower size bound or  \cite{zhu:2016} approach. 
The degree of underestimation can be substantial if the follow-up time varies greatly across patients.
The result is consistent with that for superiority trials \citep{tang:2015, tang:2017}.
We also provide theoretical justification why the quasi-Poisson regression provides quite poor control of the type I error in case of unequal follow-up, which is confirmed by simulation.

It can be challenging to specify parameters such as the event rate in the control arm and the dispersion parameter $\kappa$ at the design stage of a trial. In particular, the estimate of $\kappa$ is generally
not directly reported in the medical literature. Appendix \ref{estkappa} describes two potential ways  to back-calculate $\kappa$ based on the published point estimates of the event rates and rate ratio. 
In practice, one may also perform interim analyses of blinded data to re-estimate these nuisance parameters and revise the sample size during the mid-course of the
 trial \citep{friede:2010, tang:2015}, in which the two treatments are typically assumed to be equally effective 
 (i.e. $\lambda_1=\lambda_0$) in estimating the nuisance parameters.

The proposed formulae are derived based on the Wald CI from the NB regression. They may not be suitable for small trials. 
 \cite{aban:2008} demonstrated
via simulations that the type I error of the Wald test may be inflated if the trial size is below $50$ patients per arm. However, this may not be a particular concern in NI and equivalence trials
since the sample size in  a NI or equivalence trial is usually large. 
The inflation of the type I error is partially because  
the MLE of $\kappa$ tends to be biased  toward $0$ \citep{saha:2005} since there is no adjustment for the loss of degrees of freedom in estimating the covariate effects in the MLE procedure. 
This is analogous to the traditional analysis of variance.
In small samples, we can employ some techniques to reduce the bias in the parameter estimation, or use  more robust tests for treatment comparison. Further research will be performed
for the analysis of small  trials.

\appendix
\section{Appendix}

\subsection{Estimation of $d_g$, $\text{E}(t_{gj})$ and $\text{E}(t_{gj}^2)$ in two designs}\label{eted}
We derive expressions for $d_g$, $\text{E}(t_{gj})$ and $\text{E}(t_{gj}^2)$. Note that  $d_g$ is required for computing $n_{r}$ and $n_d$, and
$\nu_{t_g}=\text{E}(t_{gj})$ and $\text{E}(t_{gj}^2)$ are needed in calculating the sample size bounds. 
We  caluate  $d_{g_u}$ and $d_{g_l}$ based on \eqref{boundd}.
Replacing $d_g$ by $d_{g_u}$ and $d_{g_l}$ in power and sample size formulae yields the corresponding  bounds.


 In design $1$, all subjects will be followed  for $\tau_c$ years, but  may discontinue with exponential loss to follow-up (mean $\delta^{-1}$ years).  \cite{tang:2015} showed that when $\delta\neq 0$,
the overall dropout rate is $w_c=1-\exp(-\delta\tau_c)$, and
\begin{eqnarray}\label{dgdes1}
\begin{aligned}
d_g & =\text{E}\left[\frac{\lambda_g t_{gj}}{ 1+\kappa\,\lambda_g t_{gj}}\right]  =\frac{1}{\kappa}-\frac{\exp(-\delta\tau_c)}{\kappa+\kappa^2\lambda_g\tau_c}-  \int_{0}^{\tau_c}  \frac{\delta \exp(-\delta t)}{\kappa+\kappa^2 \lambda_g t}dt =\int_{0}^{\tau_c} \frac{\lambda_g \exp(-\delta t) }{(1+\kappa\lambda_g t)^2}dt,\\
\text{E}(&t_{gj})  = \frac{1-\exp(-\delta\tau_c)}{\delta} =\frac{w_c}{\delta},\\
\text{E}(&t_{gj}^2)=\frac{ 2[1-(1+ \delta\tau_c)\exp(-\delta\tau_c)]}{\delta^2}.
\end{aligned}
\end{eqnarray}
If there is no dropout (i.e. $\delta=0$), then $d_g=d_{g_l}=d_{g_u} = \frac{\lambda_g\tau_c}{ 1+\kappa\,\lambda_g \tau_c}$,  $\text{E}(t_{gj}) =\tau_c$ and $\text{E}(t_{gj}^2)=\tau_c^2$.

In design 2,  subjects are enrolled during an accrual period of $\tau_a$ years, and  followed for an additional $\tau_c$ years 
after the closure of recruitment. 
The  loss to follow-up distribution is the same as design $1$. The entry time  distribution of $e_{gj}$ is given in \eqref{entry}. 
When $\delta\neq 0$ and $\eta\neq 0$,  we get by the double expectation formula 
\begin{eqnarray}\label{dgdes2}
\begin{aligned}
\text{E}(&t_{gj})= \text{E}[\text{E}(t_{gj} | e_{gj})]= \frac{1}{\delta}- \frac{\exp(-\delta\tau_c)}{\delta} h_1,\\
 \text{E}(&t_{gj}^2)= \frac{2}{\delta^2} \left\{1- \exp(-\delta\tau_c)[(\delta\tau_c+1) h_1 +\delta h_2]\right\},\\
d_g & =\frac{1}{\kappa}-     \int_{0}^{\tau_c}  \frac{\delta \exp(-\delta t)}{\kappa+\kappa^2 \lambda_g t}dt  -  \int_{0}^{\tau_a}  \frac{\exp(-\delta_g (t+\tau_c))}{\kappa+\kappa^2\lambda (t+\tau_c)} h(t) dt
 =\int_{0}^{\tau_c+\tau_a} \frac{\lambda_g \pi(t) }{(1+\kappa\lambda_g t)^2}dt,  \\
\end{aligned}
\end{eqnarray}
where $\tau=\tau_a+\tau_c$, $\pi(t)=\exp(-\delta t)$ if $t\leq \tau_c$, $\pi(t)= \exp(-\delta t) \frac{1-\exp[-\eta(\tau-t)]}{1-\exp[-\eta\tau_a]}$ if $t>\tau_c$, and 
\begin{eqnarray}\label{dgdes3}
\begin{aligned}
h(t) & =\frac{ \delta+(\eta-\delta) \exp(\eta(t-\tau_a))}{1-\exp(-\eta\tau_a)} =   \frac{ \delta[1-\exp(-\eta(\tau_a-t))] +\eta \exp[-\eta(\tau_a-t)]}{1-\exp(-\eta\tau_a)}  ,  \\
h_1&= \frac{\eta}{\eta-\delta} \frac{\exp(-\delta\tau_a)-\exp(-\eta\tau_a)}{1-\exp(-\eta\tau_a)}  =\frac{\eta\exp(-\eta\tau_a)}{1-\exp(-\eta\tau_a)} \, \frac{1- \exp[-(\delta -\eta)\tau_a]}{\delta -\eta},\\
h_2&=\frac{\eta\exp(-\eta\tau_a)}{1-\exp(-\eta\tau_a)}\, \frac{ 1-[(\delta-\eta)\tau_a+1]\exp[-(\delta-\eta)\tau_a] }{(\delta-\eta)^2}.\\
\end{aligned}
\end{eqnarray}
When $\eta=\delta$,  $h_1$ and $h_2$ are replaced by their limiting values $h_1=\frac{\tau_a\eta\exp(-\eta\tau_a)}{1-\exp(-\eta\tau_a)}$ and $h_2=\frac{\tau_a^2\eta\exp(-\eta\tau_a)}{2[1-\exp(-\eta\tau_a)]}$.

For uniform patient entry ($\eta\rightarrow 0$) and $\delta\neq 0$, \eqref{dgdes3} reduces to 
\begin{eqnarray*}\label{dgdes4}
\begin{aligned}
h(t) & = \delta \frac{\tau_a-t}{\tau_a} +\frac{1}{\tau_a}, \\
h_1 & =\frac{1-\exp(-\delta\tau_a)}{\delta\tau_a},\\
h_2 &= \frac{1-(\delta\tau_a+1)\exp(-\delta\tau_a) }{\delta^2\tau_a},\\
\end{aligned}
\end{eqnarray*}
and $\pi(t)=\exp(-\delta t)$ if $t\leq \tau_c$, $\pi(t)=\frac{\tau-t}{\tau_a} \exp(-\delta t) $ if $t>\tau_c$.

In design $2$,  when $\delta=0$, we have $\text{E}(t_{gj})=\tau -\text{E}(e_{gj})$ and $\text{E}(t_{gj}^2)=[\tau -2\text{E}(e_{gj})] \tau+\text{E}(e_{gj}^2)$, where
 $\text{E}(e_{gj})=\tau_a/2$, $\text{E}(e_{gj}^2)=\tau_a^2/3$ if $\eta\rightarrow 0$,
and $\text{E}(e_{gj})=\frac{1-(\eta\tau_a+1)\exp(-\eta\tau_a)}{\eta [1-\exp(-\eta\tau_a)]}$, 
$\text{E}(e_{gj}^2)=\frac{2 - [\eta^2\tau_a^2 +2\eta\tau_a+2]\exp(-\eta\tau_a) }{\eta^2 [1-\exp(-\eta\tau_a)]}$
if  $\eta\neq 0$.  When $\delta=0$, $d_g$ can still be calculated using the last equality in \eqref{dgdes2}.

If the dropout rate or the dispersion parameter vary by the treatment group,  $d_g$, $\text{E}(t_{gj})$ and $\text{E}(t_{gj}^2)$ can be computed
by replacing  $\delta_g$ and/or $\kappa_g$  by the treatment specific values.

\subsection{Back-calculation of $\kappa$ based on published summary results}\label{estkappa}
{\flushleft{\bf Back-calculation of $\kappa$ based on results from the NB regression}}

Suppose a historical trial is analyzed using the NB regression. Let firstly assume the event rate for each arm is reported.
Let $\hat\lambda_g$ be the estimated event rate, and $\hat{V}_{\lambda_g}$  the associated variance for group $g$. Using \eqref{boundd} and the result in \cite{tang:2015}, we can show that
the variance $\hat{V}_{\gamma_g}=\hat{V}_{\lambda_g}/\hat\lambda_g^2$
for $\hat\gamma_g=\log(\hat\lambda_g)$  is bounded by 
\begin{equation}\label{boundk1}
 \frac{1}{n_g\hat{d}_{g_u}} \leq \hat{V}_{\gamma_g}= [\sum_{j=1}^{n_g} \frac{\hat\lambda_g t_{gj}} {1+\hat\kappa\hat\lambda_g t_{gj}}]^{-1}  \leq \frac{1}{n_g \hat{d}_{g_l}^*},
\end{equation}
where $\bar{t}_g$ and $t_{m_g}$ denote respectively  the mean and maximum follow-up time in  group $g$, and 
$$\hat{d}_{g_u} =\frac{ \hat{\lambda}_g\bar{t}_g}{1+\hat\kappa \hat{\lambda}_g\bar{t}_g} 
 \text{ and } \hat{d}_{g_l}^*=  \frac{\hat{\lambda}_g\bar{t}_g}{ 1+\hat\kappa \hat{\lambda}_g t_{m_g}}.$$
Inverting \eqref{boundk1} yields 
\begin{equation}\label{boundk11}
 n_g \hat{V}_{\gamma_g} - \frac{1}{ \hat\lambda_g  t_{m_g}} \leq  \hat\kappa \leq  n_g \hat{V}_{\gamma_g} - \frac{1}{ \hat\lambda_g \bar{t}_g}.
\end{equation}
In \eqref{boundk11}, $\hat\lambda_g \bar{t}_g$ is the expected mean number of events in group $g$, and it may be replaced by the observed mean number of events if $\bar{t}_g$ is not reported in the literature.

Now suppose the event rate ratio and its variance are reported. The variance $\hat{V}_{\gamma_{01}}$ of  $\log(\hat\lambda_1/\hat\lambda_0)=\hat\gamma_1-\hat\gamma_0$ is bounded by
\begin{equation}\label{boundk2}
 \frac{1}{n_0\hat{d}_{0_u}}+ \frac{1}{n_1\hat{d}_{1_u}} \leq \hat{V}_{\gamma_{01}} \leq  \frac{1}{n_1 \hat{d}_{1_l}^*}+ \frac{1}{n_0 \hat{d}_{0_l}^*}.
\end{equation}
Inverting \eqref{boundk2} yields
\begin{equation}\label{boundk21}
\frac{ \hat{V}_{\gamma_{01}}  -(n_0 \hat\lambda_0 \bar{t}_0)^{-1} -(n_1 \hat\lambda_1 \bar{t}_1)^{-1} }{ t_{m_0}/(n_0 \bar{t}_0)+ t_{m_1}/(n_1\bar{t}_1)} \leq \hat\kappa
    \leq  \frac{ \hat{V}_{\gamma_{01}}  -(n_0 \hat\lambda_0 \bar{t}_0)^{-1} -(n_1 \hat\lambda_1 \bar{t}_1)^{-1} }{n_0^{-1}+n_1^{-1}}.
\end{equation}

We illustrate the calculation with an example.  \cite{wang:2009} reported the analysis of a two-arm MS trial, in which $n_1=627$ was randomized to the active treatment, and $n_0=315$ subjects 
received placebo. The mean number of events was $1.1$ in the placebo arm, and $0.4$ in the active arm. 
The mean follow-up time was $1.80$ years for placebo and $1.88$ years for the active treatment. The maximum treatment duration is $2$ years.
The rate ratio and its $95\%$ CI from the NB regression is $31.3\%\, (25.2 - 38.9\%)$.
Log-transformation of the CI for $\lambda_1/\lambda_0$ yields the estimate of $\hat\gamma_1-\hat\gamma_0$ ($95\%$ CI) as $-1.162$ $(-1.378, -0.944)$, and its variance is $0.0122$. 
The event rate estimate from the NB regression was not reported for each individual treatment. We replace $\hat\lambda_g \bar{t}_g$ by
 the observed mean number of events $\bar{n}_g$, and estimate $\hat\lambda_g$ using $\bar{n}_g / \bar{t}_g$. An application of \eqref{boundk21} yields
$1.033\leq \hat\kappa \leq 1.113$, and this is roughly consistent with the reported MLE $\hat\kappa=0.99$. The  difference in the estimate of $\kappa$ arises possibly due to two main reasons.
Firstly, we do not know the MLE of $(\hat\lambda_0,\hat\lambda_1)$. Secondly, we ignore the fact that the analysis in  \cite{wang:2009} adjusted for some covariates. 
In general, inclusion of important risk factors in the model reduces the heterogeneity and $\kappa$. 

{\flushleft{\bf Back-calculation of $\kappa$ based on results from the quasi-Poisson regression}}

This method is similar to that described in \cite{zhu:2014}, but we correct an error in their formula. In the quasi-Poisson regression,
the event rate estimate is $\hat\lambda_g = \sum_{j=1}^{n_g} y_{gj}/\sum_{j=1}^{n_g} t_{gj}$  by assuming that the event counts follow the Poisson distribution.
The variance  of $\hat\lambda_g$ is firstly derived under the Poisson assumption, and  then inflated by a factor of $\phi$ to adjust for potential overdispersion \citep{wang:2009}
$$ \text{var}(\hat\lambda_g) = \phi \hat\lambda_g /(n_g \bar{t}_{g}) \text{ and }   \text{var}(\hat\gamma_g) =\text{var}(\log(\hat\lambda_g))= \phi  /(n_g \hat\lambda_g\bar{t}_{g}).$$ 
Therefore, $\phi$ can be back-calculated as
$$ \hat\phi =n_g \hat\lambda_g\bar{t}_{g}  \text{var}(\hat\gamma_g) =n_g \bar{t}_{g} \text{var}(\hat\lambda_g) / \hat\lambda_g.$$
The CI of $\hat\lambda_g$ is calculated as 
$[\exp(\hat\gamma_g -z_{1-\alpha/2} \sqrt{ \text{var}(\hat\lambda_g)}),  \exp(\hat\gamma_g +z_{1-\alpha/2} \sqrt{ \text{var}(\hat\lambda_g)})]$. 
If the CI of $\hat\lambda_g$ is reported, one needs to  log-transform the CI of $\hat\lambda_g$ to get the CI of $\hat\gamma_g$ before 
back-calculating  $\text{var}(\hat\gamma_g)$ or $\text{var}(\hat\lambda_g)$. See \cite{zhu:2014} for a numerical illustration. 
If the CI of $\lambda_1/\lambda_0$ is reported, one can derive  $\text{var}(\hat\gamma_1-\hat\gamma_0) $, 
and then back calculate $\phi$ using
$$\hat\phi = \frac{\text{var}(\hat\gamma_1-\hat\gamma_0)}{1/(n_1\lambda_1\bar{t}_1)+1/(n_0\lambda_0\bar{t}_0)}.$$

In the quasi-Poisson regression, the estimate of  $\phi$ is given by
$$\hat\phi =\frac{\sum_g\sum_j (y_{gj}-\hat{\mu}_{gj})^2/\hat\mu_{gj}}{n-p},$$
which is roughly an unbiased estimate of $1+\kappa \bar\mu$, where $p$ is the rank of covariates, and
$\bar{\mu}=\sum_g\sum_j \mu_{gj}/n$  is the expected mean
number of events among all subjects used in the analysis. 
We can estimate $\bar{\mu}$ by $p_1 \lambda_1\bar{t}_1+p_0\lambda_0\bar{t}_0$ if there is no other covariate except the treatment status. An estimate of $\kappa$ is given by
\begin{equation}\label{scaleest}
 \hat\kappa = (\hat\phi-1)/\bar\mu.
\end{equation}
 \cite{zhu:2014} suggested a wrong estimate $\hat\kappa=(\hat\phi-1)/\bar\lambda$, where $\bar\lambda$ is the pooled event rate in the two arms.

In the MS trial reported by \cite{wang:2009}, $\hat\phi=1.828$. The overall mean number of observed events is $\bar\mu=(1.1*315+0.4*627)/(315+627)=0.634$. The use of Zhu-Lakkis formula yields $\hat\kappa=2.436$, which is clearly too far from the MLE of $\kappa$ from the NB regression. 
An application of \eqref{scaleest} yields $\hat\kappa=1.306$. It provides quite a good initial guess about the dispersion $\kappa$  given the limited information.

\subsection{Explanation of  type I error inflation in quasi-Poisson regression}\label{typei}
Let consider a simple situation. We assume the same mean follow-up time ($\bar{t}_1=\bar{t}_2=\bar{t}$) and the same
sample size ($n_1=n_0=\bar{n}$)  in the two arms. Suppose the data follow the NB distribution $y_{gj} \sim \mathcal{NB}(\gamma_gt_{gj},\kappa)$, and
this  holds approximately in many empirical studies \citep{glynn:1996, wang:2009, aban:2008}.
In the quasi-Poisson regression, the event rate estimate $\hat\lambda_{pg} = \sum_{j=1}^{n_g} y_{gj}/\sum_{j=1}^{n_g} t_{gj}$ is  unbiased for $\lambda_g$.
The  variance of the log relative risk estimate $\hat\beta_p=\log(\hat\lambda_{p1}/\hat\lambda_{p0})$ can be derived by the delta method,
\begin{equation}\label{truevar}
\text{var}_{\text{true}}(\hat\beta_p)= \frac{n_1\bar{t}_1+\kappa \gamma_1 \sum_j t_{1j}^2}{n_1^2\bar{t}_1^2\gamma_1} +
           \frac{n_0\bar{t}_0+\kappa \gamma_0 \sum_j t_{0j}^2}{n_0^2\bar{t}_0^2\gamma_0}=  
   \frac{1}{n_1\bar{t}_1\gamma_1} +
           \frac{1}{n_0\bar{t}_0\gamma_0}  +\kappa [\frac{\sum_j t_{1j}^2}{\bar{n}^2\bar{t}^2} + \frac{\sum_j t_{0j}^2}{\bar{n}^2\bar{t}^2}].
\end{equation}
By the result in Appendix \ref{estkappa}, the variance estimate in the quasi-Poisson regression is given by
\begin{equation}\label{poivar}
\text{var}_{\text{poi}}(\hat\beta_p)= (1+\kappa\mu)[\frac{1}{n_1\lambda_1\bar{t}_1}+\frac{1}{n_0\lambda_0\bar{t}_0}] =   \frac{1}{n_1\bar{t}_1\gamma_1} +
           \frac{1}{n_0\bar{t}_0\gamma_0}  + \kappa \frac{(\lambda_0+\lambda_1)^2}{2\bar{n}\lambda_0\lambda_1}.
\end{equation}
Thus $$ \text{var}_{\text{true}}(\hat\beta_p) -\text{var}_{\text{poi}}(\hat\beta_p) = \frac{\kappa}{\bar{n}}\left[ \frac{\sum_j (t_{1j}-\bar{t})^2}{\bar{n}\,\bar{t}^2}  +\frac{\sum_j (t_{0j}-\bar{t})^2}{\bar{n}\,\bar{t}^2} - 
\frac{(\lambda_0-\lambda_1)^2}{2\lambda_0\lambda_1}\right] 
 \approx \frac{\kappa}{\bar{n}}\left[ 2 \text{CV}_t^2 - 
\frac{(\lambda_0-\lambda_1)^2}{2\lambda_0\lambda_1}\right], $$
where $\text{CV}_t$ is the coefficient of variation for $t_{gj}$, which is assumed  to be the same in the two arms.
Note that $(\lambda_0-\lambda_1)^2/(2\lambda_0\lambda_1)=0$ under $H_0:\lambda_0=\lambda_1$ in superiority trials, and $(\lambda_0-\lambda_1)^2/(2\lambda_0\lambda_1)\approx 0$ 
under $H_0$ in NI and equivalence trials. For example, when the margin on the rate ratio metric is $M_{r0}=1.3$ in a NI trial  
(i.e.  $\lambda_1/\lambda_0=1.3$ under $H_0$), we have
$(\lambda_0-\lambda_1)^2/(2\lambda_0\lambda_1)=0.035$. 
When the follow-up times vary greatly across patients, the quasi-Poisson approach underestimates the variance of   $\hat\beta_p$ under $H_0$,  leading to the
type I error inflation.

\subsection{Technical proofs}\label{proof}

\begin{proof}{\bf of  \eqref{equiv}:}
Let $\bar\lambda=\sqrt{\lambda_0\lambda_1}$. Then $\lambda_0=\bar{\lambda}\exp(-\beta/2)$, $\lambda_1=\bar{\lambda}\exp(\beta/2)$ and 
$\lambda_0-\lambda_1=\bar{\lambda}[\exp(-\beta/2) - \exp(\beta/2)]\approx -\bar{\lambda} \beta$.
We can approximate \eqref{sizesnon}  by $\tilde{n}_r=[ ( p_0\tilde{d})^{-1}+(p_1\tilde{d})^{-1}] f$, and \eqref{sizenondiff}
by $\tilde{n}_d= [\bar\lambda^2/(p_0\tilde{d})+\bar\lambda^2/(p_1\tilde{d})] (z_{1-\alpha/2}+z_P)^2/(M_{d0}+\bar{\lambda} \beta)^2$,
where $\tilde{d} = E[\kappa \bar\lambda/(1+\kappa\bar\lambda)]$.
If \eqref{equiv} holds, then $n_d\approx \tilde{n}_d = \tilde{n}_r\approx n_r$.
\end{proof}

\begin{proof}{\bf of  \eqref{sizeequiv2}:}  The power in \eqref{powerequiv} satisfies
$$P +1 = \Phi\left(\frac{\sqrt{n}[\log(M_{ru})-\beta]}{\sqrt{ (d_0p_0)^{-1}+(d_1p_1)^{-1}}}-z_{1-\alpha/2}\right) +
         \Phi\left(\frac{\sqrt{n}[\beta-\log(M_{rl})]}{\sqrt{ (d_0p_0)^{-1}+(d_1p_1)^{-1}}}-z_{1-\alpha/2}\right)$$
which is bounded by
\begin{equation*}
2 \Phi\left(\frac{\sqrt{n} \Delta_{min}}{\sqrt{ (d_0p_0)^{-1}+(d_1p_1)^{-1}}}-z_{1-\alpha/2}\right) \leq P+1\leq 2\Phi\left(\frac{\sqrt{n}\Delta_{max}}{\sqrt{ (d_0p_0)^{-1}+(d_1p_1)^{-1}}}-z_{1-\alpha/2}\right).
\end{equation*}
Inverting the above inequalities yields \eqref{sizeequiv2}.
\end{proof}

{\flushleft{\bf Online Supplementary Materials}}\\
The online supplementary materials contain the SAS macro for calculating the power and sample size for comparing NB rates based on both the absolute and relative rate difference in
 superiority, NI and equivalence trials. It allows the dropout rate and dispersion parameter to vary by treatment group. A R package with similar functions is in development.

SAS MACRO is provided for implementing the  power and sample size calculation for comparing NB rates based on both the absolute and relative rate differences in superiority, NI and equivalence
trials. It allows the dropout rate and dispersion parameter to vary by treatment group. A R package with
similar functions is in development. 

Below are key equations and notations used in the macro.
Let $Y\sim \mathcal{NB}(\mu,\kappa)$ denote the  Negative Binomial (NB) distribution with mean $\mu$, dispersion $\kappa$ and variance $\mu+\kappa \mu^2$.
Suppose in a clinical trial, $n$ subjects are randomized to the experimental ($g=1$) or control ($g=0$) treatment.
Let $t_{gj}$ be  the follow-up time, and $y_{gj} \sim \mathcal{NB}(\mu_{gj}=\lambda_g t_{gj},\kappa_g)$  the number of observed events for subject $j=1,\ldots,n_g$ in group $g$. 
  
For the rate ratio metric, we let $p_g=n_g/n$, $d_g=\text{E}[\lambda_g t_{gj}/(1+\kappa_g\lambda_g t_{gj})]$,
$$ \sigma^2=\frac{1}{ d_0p_0}+\frac{1}{d_1p_1}, \beta=\log(\lambda_1/\lambda_0), \beta^*=\log(M_{r0})-\beta,$$
$$\Delta_{a}=\log(M_{ru})-\beta, \Delta_b=\log(M_{rl})-\beta, \Delta_{min}=\min[\Delta_a, -\Delta_b], \Delta_{max}=\max[\Delta_a, -\Delta_b]$$
where $M_{r0}$ is the NI margin, $(M_{rl}<1, M_{ru}>1)$ is the equivalence margin.

For the absolute rate difference metric, we define
$$ \sigma^2 =\frac{ \lambda_0^2}{d_0p_0}+ \frac{\lambda_1^2}{d_1p_1}, \beta= \lambda_1 - \lambda_0,  \beta^*= M_{d0}-\beta, $$
$$ \Delta_a=M_{du}-\beta, \Delta_b= M_{dl}-\beta,  \Delta_{min}=\min[\Delta_a, -\Delta_b], \Delta_{max}=\max[\Delta_a, -\Delta_b]$$
where $M_{d0}$ is the NI margin, $(M_{dl}<0, M_{du}>0)$ is the equivalence margin.

The power and sample size formulae for a NI trial are given by
\begin{eqnarray*}
\begin{aligned}
P = \Phi\left(\frac{\sqrt{n}|\beta^*|}{\sqrt{\sigma^2}}-z_{1-\alpha/2}\right) \text{ and } n =  \frac{(z_{1-\alpha/2}+z_{P})^2\sigma^2}{\beta^{*^2}}.
\end{aligned}
\end{eqnarray*}
The formulae for a superiority trial can be obtained by setting $M_{r0}=1$ on the rate ratio metric or $M_{d0}=0$ on the absolute rate difference metric.
The formulae for NI trials are valid if \\
(1) $M_{r0}>1$, $\lambda_1/\lambda_0<M_{r0}$: to show that the active rate is not materially higher than the control rate.\\
(2) $M_{r0}<1$, $\lambda_1/\lambda_0>M_{r0}$: to show that the active rate is not materially lower than the control rate.\\
(3) $M_{d0}>0$, $\lambda_1-\lambda_0<M_{d0}$: to show that the active rate is not materially higher than the control rate.\\
(4) $M_{d0}<0$, $\lambda_1-\lambda_0>M_{d0}$:  to show that the active rate is not materially lower than the control rate.\\

In an equivalence trial,  the power is  
\begin{eqnarray*}\label{powerequiv}
\begin{aligned}
P = \Phi\left(\frac{\sqrt{n}\Delta_{a}}{\sqrt{\sigma^2}}-z_{1-\alpha/2}\right) -
         \Phi\left(\frac{\sqrt{n}\Delta_{b}}{\sqrt{ \sigma^2}}+z_{1-\alpha/2}\right) =\Phi\left(\frac{\sqrt{n}\Delta_{a}}{\sqrt{ \sigma^2}}-z_{1-\alpha/2}\right) +
         \Phi\left(\frac{-\sqrt{n}\Delta_{b}}{\sqrt{ \sigma^2}}-z_{1-\alpha/2}\right)-1 .
\end{aligned}
\end{eqnarray*}
If the sample size is too small, $P$ can be negative, which will be set to $P=0$.
The sample size is bounded by 
$$\frac{(z_{1-\alpha/2}+z_{(1+P)/2})^2\sigma^2}{\Delta_{max}^2} \leq n \leq \frac{(z_{1-\alpha/2}+z_{(1+P)/2})^2\sigma^2}{\Delta_{min}^2}.$$
If $\Delta_{min}=\Delta_{max}$, both the lower and upper bounds are equal to the desired sample size. Otherwise the bisection method is used to calculate the sample size,
at which the power is no smaller than the target power.
In the equivalence trial, we require that $\lambda_1/\lambda_0$ lies within the interval $[M_{rl},M_{ru}]$ (rate ratio metric) or that $\lambda_1 - \lambda_0$ lies within the interval $[M_{dl},M_{du}]$ (absolute rate
difference metric).
That is $\Delta_{a}>0$, $\Delta_b<0$, and $\Delta_{max}\geq \Delta_{min}>0$.

Note that $d_g$ is bounded by 
$$d_{g_l} = \frac{\lambda_g \text{E}^2(t_{gj})}{\text{E}(t_{gj})+\kappa_g  \lambda_g \text{E}(t_{gj}^2)}\leq d_g \leq d_{g_u} =  \frac{\lambda_g \text{E}(t_{gj})}{1+\kappa_g \lambda_g\text{E}(t_{gj})},$$
Replacing $d_g$ by $d_{g_u}$ and $d_{g_l}$ in the power and sample size formulae yields the corresponding lower and upper bounds on power and sample size.
We calculate $d_g$,  $d_{g_u}$, $d_{g_l}$, $\text{E}(t_{gj})$ and $\text{E}(t_{gj}^2)$  using the formulae in Appendix A.1 of the main paper.

\begin{verbatim}
*** Power & sample size for comparing Negative Binominal rates;
*** in superiority noninferiority & equivalence trials with dropouts;
*** The methods allow the dispersion parameter and dropout rates to
    vary by treatment group, and the treatment comparison can be based 
    on both the absolute and relative rate difference metrics;
*** If you have any comments & suggestions, please send them to       
    yongqiang_tang@yahoo.com;
*** Reference        
    1. Tang, Y., 2015. Sample size estimation for negative binomial    
       regression comparing rates of recurrent events with unequal 
       follow-up time. 
       Journal of Biopharmaceutical Statistics 25, 1100-13; 
*** 2. Tang Y. 2017, Sample size for comparing negative binomial rates    
       in noninferiority and equivalence trials with unequal follow-up 
       time. Journal of Biopharmaceutical Statistics (accepted);
*** 3. Tang, Y., 2017. Negative binomial regression: Sample size with  
       unequal follow-up times. In: Encyclopedia of Biopharmaceutical 
       Statistics, Third Edition DOI: 10.1081/E-EBS3-140000049;
       
*** The estimate in Zhu (2017, Journal of Biopharmaceutical research, 
    vol. 9, page 107-115) or Zhu-Lakkis (2014, statistics in Medicine, 
    vol. 33, page 376â€“387) is produced for rate ratio based test if 
    the diserpsion and dropout rates do not vary with treatment group;
*** Zhu's estimate is generally close to Tang's lower size bound; 
    
%macro NBsize(lambda0=1,lambda1=0.8,kappa0=1,kappa1=,tauc=2,taua=2,
droprate0=0.2,droprate1=,p0=0.5,alpha=0.05,power=0.8,ntot=,Mr0=1.3,
Mru=1.3, Mrl=, Md0= 0.262, Mdu=0.262, Mdl=, eta=0, type =SUP,
metric=ratio, design=1);
*** lambda0,lambda1 -- event rates in control and active arms ;
*** kappa0,kappa1--dispersion parameter for the control & active arms;
    **if kappa1=., it will be set to kappa1=kappa0 (same in two arms);
*** droprate0, droprate1 --- exponential dropout rate in two arms;    
    **if droprate1=., it will be set to droprate1=droprate0;
*** Mr0 -- noninferiority (NI) margin on rate ratio;
*** Mru, Mrl --- Equivalence margins on rate ratio;
    ** If MRl=., it will be set to Mrl=1/Mru;
*** Md0 -- noninferiority (NI) margin on absolute rate difference;
*** Mdu, Mdl --- Equivalence margins on absolute rate difference; 
    ** if Mdl=. , it will be set to Mdl=-Mdu;
*** Metric = RATIO (treatment is compared on the rate ratio metric);
    ** Metric=DIFF (absolute rate difference treatment effect metric);
*** In design=1, the planned treatment duration is tauc time unit for 
     all subjects;
*** In design=2, the enrollment period is taua Time Unit. Subject will  
     be followed for an additional tauc UNIT after the recruitment 
     period. The total study duration is tau=tauc+taua;
*** In design=2, the entry time distribution is uniform if eta=0, 
    eta * exp(-eta*t)/[1-exp(-eta *taua)] if eta^=0;
*** Ntot -- total sample size, if Ntot is not missing, nominal power 
    will be calculated at the given Ntot;
*** Power- target power, if not missing, sample size will be computed;
   
   data _NULL_;
   if &droprate1=. then call symput('droprate1',&droprate0);
   if &kappa1=.  then call symput('kappa1',&kappa0);
   if &mrl=. then call symput('mrl',1/&mru);
   if &Mdl=. then call symput('Mdl',-&Mdu);
   run;
   
   data parametercheck;
   length error $200.;
   nerror=0;
   if (upcase("&metric") in ('RATIO', 'DIFF'))^=1 then do;
      nerror=nerror+1;
      error='Metric should be equal to RATIO or DIFF';
   end;   
   else if (upcase("&type") in ('SUP','NI','EQUI'))^=1 then do;
      nerror=nerror+1;
      error='Type should be equal to SUP, NI or EQUI';
   end; 
   else if (&droprate1<0)  | (&droprate0 <0) | (&lambda1<=0) | 
           (&lambda0<=0) | (&kappa0<0) | (&kappa1<0)  | (&tauc<=0) 
   then do;
      nerror=nerror+1;
    error='Error: droprate0/droprate1,lambda0/lambda1,kappa0/kappa1,
    tauc shall be non-negative or positive';
   end;
   else if ("&design" in ('1','2'))^=1  then do;
     nerror=nerror+1;
    error= 'Error: design should be equal to 1 or 2';
  end;
  else if ((&power^=.) & (&power<=0)) | (&power>=1) | 
          ((&ntot^=.) & (&ntot<0)) | ((&power=.) & (&ntot=.)) | 
       ((&power>0) & (&ntot>0))  then do;
     nerror=nerror+1;
    error= 'Error: there should be either 0<power<1, ntot=. 
              OR ntot>0 & power=.';
  end;
   else if &design=2 & &taua<0 then do;
     nerror=nerror+1;
    error= 'Error: taua must be >0 in design 2';
  end;
  else if &p0<0 | &p0>1  then do; 
   nerror=nerror+1;
  error='Error: p0 the proprotion of subject in control
       arm must be between 0 and 1';
  end;
  else if index(upcase("&type"), 'SUP')>0 & &lambda1=&lambda0 
    then do;
   nerror=nerror+1;
   Error='Error: lambda0 & lambda1 should be different in a 
          superiority trial';
  end;
  else if index(upcase("&type"),'EQUI')>0 & upcase("&metric")='DIFF' &
        ((&Mdu=.) | (&Mdl=.)| (&lambda1-&lambda0>=&Mdu) | 
        (&lambda1-&lambda0<=&Mdl)) then do;
      nerror=nerror+1;
    error='Error: Equivalence DIFF margin must satisfy
         Mdl< lambda1-lambda0<Mdu, Mdu^=.';
  end;
  else if index(upcase("&type"),'EQUI')>0 & upcase("&metric")='RATIO' &
        ((&lambda1/&lambda0>=&Mru) | (&lambda1/&lambda0<=&Mrl)) then do;
      nerror=nerror+1;
    error='Error: Equivalence RATIO margin must satisfy 
       Mrl< lambda1/lambda0<Mru, Mru^=.';
  end;
  else if index(upcase("&type"),'NI')>0 & upcase("&metric")='RATIO' &
        ((&MR0=.)| (&Mr0<=0) | (&lambda1/&lambda0=&Mr0)) then do;
         nerror=nerror+1;
    error='Error: NI RATIO Margin must satisfy Mr0>0 & lambda1/lambda0^= Mr0';
  end;
  else if index(upcase("&type"),'NI')>0 & upcase("&metric")='DIFF' &
      ((&Md0=.) |  (&lambda1-&lambda0=&Md0)) then do;
       nerror=nerror+1;
    error='Error: NI DIFF margin must satisfy MD0^=. and lambda1-lambda0^= Md0';
  end;
  call symput('nerror', nerror);
   run;
 %if &nerror>0 %then %do;
  proc print data=parametercheck noobs;var error;run;
 %end;
%if &nerror<=0 %then %do;
  proc iml;  
    start d0fordesign1(design) global(lambda, kappa, droprate,tauc);
        start d0fun(t) global(droprate, lambda, kappa);
              vv =exp(- droprate *t) *lambda /((1+kappa*lambda*t)**2);
           return(vv); 
        finish; 
        if droprate^=0 then do;
           c  = 0||tauc; 
           call quad(d0,"d0fun",c); 
           tbar = (1-exp(-droprate *tauc))/droprate;
           t2bar = 2*(1-(droprate *tauc+1)*exp(-droprate*tauc))/(droprate**2);
           du = lambda*tbar/(1+kappa*lambda*tbar);
           dl = lambda*(tbar**2)/(tbar+kappa *lambda*t2bar);
         end; 
         if droprate =0 then do;
           tbar=tauc;
           t2bar =tauc**2;
           d0= lambda*tauc/(1+kappa*lambda*tauc);
           du=d0;
           dl=d0;
         end;
         return (d0||dl||du||tbar||t2bar);
    finish;
    
    start d0fordesign2(design) global(lambda, kappa, droprate,eta, taua, tauc);
        start d0fun2(t) global(droprate, lambda, kappa, eta, taua,tauc);
              vv =exp(- droprate *t) *lambda /((1+kappa*lambda*t)**2);
              if t>tauc then do;
               if eta^=0 then vv=vv*(1-exp(-eta*(tauc+taua-t)))/(1-exp(-eta*taua));
               else if eta=0 then vv=vv* (tauc+taua-t)/taua;
              end;
           return(vv); 
        finish;      
         c  = 0||(tauc+taua); 
         call quad(d0,"d0fun2",c);
        if droprate^=0 then do;
                if eta^=0 then  temp = eta * exp(-eta *taua)/(1-exp(-eta*taua));
                else temp= 1/taua;          
                if droprate=eta then do;
                       hg1=temp*taua;
                       hg2=taua**2/2*temp;
                 end;
                 else do;
   hg1 = (1-exp(-(droprate-eta)*taua))/(droprate-eta)*temp;
  hg2=(1-((droprate-eta)*taua+1)*exp(-(droprate-eta)*taua))/((droprate-eta)**2)*temp;
                 end;        
        tbar=(1 -exp(-droprate * tauc)*hg1)/droprate;
  t2bar = 2*( 1-((droprate*tauc+1)*hg1+droprate *hg2)*exp(-droprate*tauc))/(droprate**2);
           end; 

          if droprate =0 then do;
             if eta=0 then do;  
                      tbar=tauc+taua/2;
                      t2bar =(tauc+taua)*tauc+taua**2/3;
              end;
              else do;
                 eattaua=eta*taua;
                 Ee = (1-(eattaua+1)*exp(-eattaua))/(1-exp(-eattaua))/eta;
                 Ee2 = (2- (eattaua**2+2*eattaua+2)*exp(-eattaua))/(1-exp(-eattaua))/(eta**2);
                 tbar=tauc+taua-Ee;
                 t2bar=(tauc+taua)**2- 2*Ee*(tauc+taua)+Ee2;
              end;
         end;   
           du = lambda*tbar/(1+kappa*lambda*tbar);
           dl = lambda*(tbar**2)/(tbar+kappa *lambda*t2bar);
         return ((d0||dl||du||tbar||t2bar));
    finish;
    
    **** estimate size for equivalence trial using bisection method;
    start size_equiv(beta, margl, margu, var, alpha, power,tolerence);       
         tempu= margu-beta;
         templ= -margl+beta;
         n_up = (probit(1-&alpha/2)+probit((&power+1)/2))**2*var/ min(tempu**2, templ**2);
         if abs(tempu-templ)<0.000000000001 then nnreq=n_up;
         else do;
            n_low= (probit(1-&alpha/2)+probit(&power))**2*var/max(tempu**2,templ**2);       
           do while (abs(n_up-n_low)>tolerence);
             n_mid=(n_low+n_up)/2;
             pow_up = probnorm(sqrt(n_mid/var)*(margu -beta) - probit(1-&alpha/2))-
                     probnorm(sqrt(n_mid/var)*(margl-beta) + probit(1-&alpha/2));
             if pow_up > power then n_up =  n_mid;
             else n_low= n_mid;
           end;      
           nnreq=n_mid;
         end;  
         return(nnreq);
    finish;
    
   start varnull(type, Mr0, p0, p1, lambda0,lambda1, kappa, ET);
      if upcase(type)="SUP" then do;
        r0tilde =p0*lambda0+p1*lambda1;
        r1tilde =r0tilde ;
      end;
      else do;
          theta=p1/p0;
          a=-kappa * et * Mr0*(1+theta);
          b=kappa*et*(lambda0*Mr0+theta* lambda1) -(1+theta* Mr0);
          c=lambda0+theta* lambda1;
         if kappa^=0 then  r0tilde = (-b-sqrt(b*b-4*a*c))/(2*a);
         if kappa=0 then  r0tilde=(lambda0+theta*lambda1)/(1+theta*Mr0);
         r1tilde=r0tilde* Mr0; 
      end;       
       v0=kappa/(p0*p1) + (1/(p0*r0tilde)+1/(p1*r1tilde))/et;
       return (v0);
   finish;
      
    if &design=1 then do;
    *** the planned duration is tauc for all subjects, exponential dropout rate;
      lambda=&lambda0; kappa=&kappa0;droprate=&droprate0;tauc=&tauc;
      par0= d0fordesign1(1);  
      droprate=&droprate1;  kappa=&kappa1;  lambda=&lambda1;
      par1= d0fordesign1(1);
    end;
    if &design=2 then do;
    *** the planned duration is tauc for all subjects, exponential dropout rate;
      lambda=&lambda0; kappa=&kappa0;droprate=&droprate0;tauc=&tauc;taua=&taua;
     eta=&eta;
      par0= d0fordesign2(2);  
      droprate=&droprate1; kappa=&kappa1; lambda=&lambda1;
      par1= d0fordesign2(2);
    end;
        
    p0=&p0; p1=1-p0;
    print  'd0 dl du E(t) E(t*t) in control arm' par0;
    print  'd0 dl du E(t) E(t*t) in active arm' par1;

if upcase("&metric") = 'RATIO' then do;
    vard0 = 1/(par0[1]*p0)+1/(par1[1]*p1);
    vardl = 1/(par0[2]*p0)+1/(par1[2]*p1);
    vardu = 1/(par0[3]*p0)+1/(par1[3]*p1);
    beta=log((&lambda1)/(&lambda0));
    if upcase("&type")='NI' then beta = log(&Mr0)-beta;
    else if upcase("&type")='EQUI' then do;
      margl=log(&mrl); margu=log(&mru);
    end;
 end;
 else do;
    vard0 = (&lambda0)**2/(par0[1]*p0)+(&lambda1)**2/(par1[1]*p1);
    vardl = (&lambda0)**2/(par0[2]*p0)+(&lambda1)**2/(par1[2]*p1);
    vardu =(&lambda0)**2/(par0[3]*p0)+(&lambda1)**2/(par1[3]*p1);
    beta=&lambda1-&lambda0;
   if upcase("&type")='NI' then  beta = &MD0 - beta; 
    else if upcase("&type")='EQUI' then do;
      margl=&Mdl; margu=&Mdu;
    end;   
 end;

    if (upcase("&type")='NI' | upcase("&type")='SUP') then do;
         if &power>0 then do;
         temp = (probit(1-&alpha/2)+probit(&power))**2/(beta**2);
         nreq_raw = temp*vard0; nreq=ceil(nreq_raw);
         nlow = ceil(temp *vardu);
         nup =  ceil(temp*vardl);
     print 'The required size (before rounding) is' nreq_raw, 'which is rounded up to' nreq;
     print 'The low and upper sample size bounds are' nlow nup; 
         power_nominal = probnorm( sqrt(nreq/vard0)*abs(beta)-probit(1-&alpha/2));
      print 'The nominal power is'  power_nominal 'at the sample size' nreq;
     
  if (&droprate1=&droprate0) & (&kappa1=&kappa0) & upcase("&metric") ='RATIO' then do; 
        v0= varnull("&type", &Mr0, p0, p1, &lambda0,&lambda1, &kappa0, par0[4]);
        nzl = probit(&alpha/2)*sqrt(v0)+ probit(1-&power)*sqrt(vardu);
        nzl=ceil(nzl**2/beta/beta);    
        print 'Zhu (2017) or Zhu-Lakkis (2014) estimate using variance under null via MLE' nzl;
         print 'It is generally close to Tang lower size bound.';
        end;
    end;
   
      if &ntot>0 then do;
            power_nominal = probnorm( sqrt(&ntot/vard0)*abs(beta)-probit(1-&alpha/2));
        print 'The nominal power is'  power_nominal 'at the sample size' (&ntot);
      end;  
  end;
   
   if upcase("&type")='EQUI' then do;
        if &power>0 then do;
         nreq_raw = size_equiv(beta, margl, margu, vard0, &alpha, &power,0.00001);
         nreq=ceil(nreq_raw);
         power_nominal  = probnorm(sqrt(nreq/vard0)*(margu-beta) - probit(1-&alpha/2))-
                     probnorm(sqrt(nreq/vard0)*(margl-beta) + probit(1-&alpha/2));
         nlow = size_equiv(beta, margl, margu, vardu, &alpha, &power,0.00001);
         nup = size_equiv(beta, margl, margu, vardl, &alpha, &power,0.00001);
         nlow=ceil(nlow);
         nup=ceil(nup);
         print 'The required size (before rounding) is' nreq_raw, 
                     'which is rounded up to' nreq 'at margin' (&Mrl) (&Mru);
         print 'The nominal power is'  power_nominal 'at the sample size' nreq;
         print 'The low and upper sample size bounds are' nlow nup;
                
  if  (&droprate1 = &droprate0) & (&kappa1=&kappa0) &  upcase("&metric")='RATIO' then do;
      v0plus= varnull("&type", &Mru, p0, p1, &lambda0,&lambda1, &kappa0, par0[4]);
      v0minus=varnull("&type", &Mrl, p0, p1, &lambda0,&lambda1, &kappa0, par0[4]);
     
      nzl=probit(&alpha/2)*sqrt(v0plus)+ probit(1-&power)*sqrt(vardu);
      nzl=ceil(nzl**2/((log(&Mru*&lambda0/&lambda1))**2));   
      prob=0;
     do while (prob<&power);
        nzl=nzl+1;  
      prob1= sqrt(nzl)*abs(log(&mru*&lambda1/&lambda0)) -probit(1-&alpha/2)*sqrt(v0minus);
      prob2 = sqrt(nzl)*abs(log(&mrl*&lambda1/&lambda0)) - probit(1-&alpha/2)*sqrt(v0plus);
        prob=probnorm(prob1/sqrt(vardu))+probnorm(prob2/sqrt(vardu))-1;
       end;
      print 'Zhu (2017) estimate using variance under null via MLE' nzl;
      print 'It is generally close to Tang lower size bound.';
    end;
  end;
  if &ntot>0 then do;
     power_nominal  = probnorm(sqrt(&ntot/vard0)*(margu-beta) - probit(1-&alpha/2))-
                     probnorm(sqrt(&ntot/vard0)*(margl-beta) + probit(1-&alpha/2));
      if   power_nominal<0 then  power_nominal  =0;
    print 'The nominal power is'  power_nominal 'at the sample size' (&ntot);
   end;
 end; 
quit;
%end;
%mend;
   
**** example 1: superiority trial, equal dropout and dispersion, to calculate sample size;
%NBsize(lambda0=0.6, lambda1=0.3, kappa0=1, kappa1=., tauc=2, taua=2, droprate0=0.178, 
droprate1=., p0=0.5, alpha=0.05, power=0.8, ntot=., Mr0=1.2, Mru=1.3, Mrl=., Md0= 0.1613, 
Mdu=0.1613, Mdl=., eta=0, type =sup, metric=ratio, design=1);
      
**** example 2: noninferiority trial, equal dropout and dispersion, to calculate sample size;
%NBsize(lambda0=0.6, lambda1=0.36, kappa0=1, kappa1=., tauc=2, taua=2, droprate0=0.3, 
droprate1=., p0=0.5, alpha=0.05, power=0.8, ntot=., Mr0=1.2, Mru=1.3, Mrl=., Md0= 0.1613, 
Mdu=0.1613, Mdl=., eta=0, type =ni, metric=ratio, design=2);    
      
**** example 3: noninferiority trial, dispersion vary by treatment, rate ratio test;
%NBsize(lambda0=0.6, lambda1=0.48, kappa0=2, kappa1=1, tauc=2, taua=2, droprate0=0.1438, 
droprate1=., p0=0.5, alpha=0.05, power=0.8, ntot=., Mr0=1.3, Mru=1.3, Mrl=., Md0= 0.1613, 
Mdu=0.1613, Mdl=.,eta=0, type =ni, metric=ratio, design=1);      
      
**** example 4: noninferiority trial, dispersion vary by treatment, rate difference metric;
%NBsize(lambda0=0.6, lambda1=0.48, kappa0=2, kappa1=1, tauc=2, taua=2, droprate0=0.1438, 
droprate1=., p0=0.5, alpha=0.05, power=0.8, ntot=., Mr0=1.3, Mru=1.3, Mrl=., Md0= 0.1408,
 Mdu=0.1613, Mdl=., eta=0, type =ni, metric=diff, design=1);  

**** example 5: equivalence trial, equal dropout and dispersion, rate ratio metric;
%NBsize(lambda0=0.6, lambda1=0.63, kappa0=1, kappa1=1, tauc=2, taua=2, droprate0=0.2, 
droprate1=., p0=0.5,alpha=0.05, power=0.8, ntot=., Mr0=1.3, Mru=1.3, Mrl=., Md0= 0.1408,
 Mdu=0.1613, Mdl=.,eta=0, type =equi, metric=ratio, design=2); 

**** example 6: equivalence trial, equal dropout and dispersion, to calculate power;
%NBsize(lambda0=0.6, lambda1=0.63, kappa0=1, kappa1=1, tauc=2, taua=2, droprate0=0.2, 
droprate1=., p0=0.5, alpha=0.05, power=., ntot=1000, Mr0=1.3, Mru=1.3, Mrl=., Md0= 0.1408,
 Mdu=0.1613, Mdl=., eta=0, type =equi, metric=ratio, design=2); 
\end{verbatim}

 \bibliographystyle{elsart-harv}
\bibliography{nbsizenon} 
\newpage

  \begin{table}[h]
\begin{center}
\begin{tabular}{ccccr cc c  c rrcccc} \\\hline  
                         &   &&\multicolumn{4}{c}{test based on rate ratio}   &&    \multicolumn{4}{c}{test based on rate difference}  \\\cline{4-7} \cline{9-12}
                       &  & & & & \multicolumn{2}{c}{type I error  ($\%$) }  &  &&&  \multicolumn{2}{c}{type I error ($\%$)}   \\\cline{6-7}\cline{11-12}
$\lambda_0$ & $\exp(\beta)$ & $\kappa$ & $M_{r0}$ & $n_{r}$ & NB & Poisson   && $M_{d0}$ &  $n_{d}$ & NB & Poisson  \\\hline 
            $0.6$& $0.65$ &$1.0$&$1.2$&$  192$&$ 2.61$&$ 2.87$&&$0.0882$&$  198$&$ 2.65$&$ 2.95$\\
            $0.6$& $0.80$ &$1.0$&$1.2$&$  412$&$ 2.49$&$ 2.79$&&$0.0978$&$  416$&$ 2.59$&$ 2.92$\\
            $0.6$&$0.95$ & $1.0$&$1.2$&$ 1185$&$ 2.47$&$ 2.70$&&$0.1066$&$ 1186$&$ 2.53$&$ 2.82$\\
            $0.6$&$1.00$ & $1.0$&$1.2$&$ 1921$&$ 2.46$&$ 2.81$&&$0.1094$&$ 1921$&$ 2.73$&$ 3.05$\\
            $0.6$&$1.05$ & $1.0$&$1.2$&$ 3540$&$ 2.37$&$ 2.62$&&$0.1121$&$ 3543$&$ 2.54$&$ 2.83$\\
            $0.6$&$0.65$ & $1.0$&$1.3$&$  150$&$ 2.80$&$ 3.05$&&$0.1269$&$  155$&$ 2.97$&$ 3.36$\\
            $0.6$&$0.80$ & $1.0$&$1.3$&$  288$&$ 2.66$&$ 3.03$&&$0.1408$&$  291$&$ 2.70$&$ 3.08$\\
            $0.6$&$0.95$ & $1.0$&$1.3$&$  658$&$ 2.48$&$ 2.76$&&$0.1534$&$  658$&$ 2.83$&$ 3.09$\\
            $0.6$&$1.00$ & $1.0$&$1.3$&$  928$&$ 2.69$&$ 2.98$&&$0.1574$&$  928$&$ 2.68$&$ 3.00$\\
            $0.6$&$1.05$ & $1.0$&$1.3$&$ 1384$&$ 2.48$&$ 2.77$&&$0.1613$&$ 1385$&$ 2.55$&$ 2.91$\\
            $0.9$&$0.65$ & $1.5$&$1.2$&$  202$&$ 2.55$&$ 2.98$&&$0.1323$&$  212$&$ 2.70$&$ 3.32$\\
            $0.9$&$0.80$ & $1.5$&$1.2$&$  442$&$ 2.63$&$ 3.08$&&$0.1468$&$  449$&$ 2.61$&$ 3.04$\\
            $0.9$&$0.95$ & $1.5$&$1.2$&$ 1294$&$ 2.51$&$ 2.94$&&$0.1599$&$ 1295$&$ 2.64$&$ 3.04$\\
            $0.9$&$1.00$& $1.5$&$1.2$&$ 2107$&$ 2.46$&$ 2.88$&&$0.1641$&$ 2107$&$ 2.54$&$ 2.89$\\
            $0.9$&$1.05$ & $1.5$&$1.2$&$ 3900$&$ 2.41$&$ 2.87$&&$0.1681$&$ 3904$&$ 2.49$&$ 2.92$\\
            $0.9$&$0.65$ & $1.5$&$1.3$&$  158$&$ 2.75$&$ 3.22$&&$0.1904$&$  166$&$ 2.78$&$ 3.39$\\
            $0.9$&$0.80$ & $1.5$&$1.3$&$  309$&$ 2.58$&$ 3.00$&&$0.2112$&$  313$&$ 2.79$&$ 3.28$\\
            $0.9$&$0.95$ & $1.5$&$1.3$&$  718$&$ 2.68$&$ 3.12$&&$0.2301$&$  719$&$ 2.80$&$ 3.28$\\
            $0.9$&$1.00$ & $1.5$&$1.3$&$ 1018$&$ 2.48$&$ 2.80$&&$0.2361$&$ 1018$&$ 2.84$&$ 3.26$\\
            $0.9$&$1.05$ & $1.5$&$1.3$&$ 1525$&$ 2.60$&$ 3.00$&&$0.2420$&$ 1526$&$ 2.78$&$ 3.14$\\
\hline
 \end{tabular} \caption{Empirical estimate of the one-sided type I error at the nominal level of $2.5\%$ based on $10000$ simulated NI trials for design 1: a)  $\tau_c=2$,  and the overall dropout rate is $25\%$ ($\delta=0.1438$);
b) $\exp(\beta)$ is used only in the sample size calculation; c) the data are simulated under the assumption that $\lambda_1=\lambda_0M_{r0}$ for the test based on the rate ratio metric, and $\lambda_1=\lambda_0+M_{d0}$ for the test based on the rate difference metric.
 }\label{type1_sim1}
\end{center}
\end{table}

  \begin{table}[h]
\begin{center}
\begin{tabular}{ccccr cc c  c rrcccc} \\\hline  
                        &    &&\multicolumn{4}{c}{rate ratio based test}   &&    \multicolumn{4}{c}{rate difference based test}  \\\cline{4-7} \cline{9-12}
                        & & & & & \multicolumn{2}{c}{type I error  ($\%$) }  &  &&&  \multicolumn{2}{c}{type I error ($\%$)}   \\\cline{6-7}\cline{11-12}
$\lambda_0$& $\exp(\beta)$  & $\kappa$ & $M_{r0}$ & $n_{r}$ & NB & Poisson   && $M_{d0}$ &  $n_{d}$ & NB & Poisson  \\\hline 
            $0.6$& $0.65$ &$1.0$&$1.2$&$  176$&$ 2.81$&$ 3.59$&&$0.0882$&$  183$&$ 2.72$&$ 3.55$\\
            $0.6$& $0.80$ &$1.0$&$1.2$&$  381$&$ 2.52$&$ 3.40$&&$0.0978$&$  385$&$ 2.60$&$ 3.39$\\
            $0.6$& $0.95$ &$1.0$&$1.2$&$ 1102$&$ 2.43$&$ 3.25$&&$0.1066$&$ 1103$&$ 2.61$&$ 3.48$\\
            $0.6$& $1.00$ &$1.0$&$1.2$&$ 1789$&$ 2.61$&$ 3.44$&&$0.1094$&$ 1789$&$ 2.60$&$ 3.50$\\
            $0.6$& $1.05$ &$1.0$&$1.2$&$ 3302$&$ 2.50$&$ 3.32$&&$0.1121$&$ 3304$&$ 2.47$&$ 3.30$\\
            $0.6$& $0.65$ &$1.0$&$1.3$&$  138$&$ 2.81$&$ 3.68$&&$0.1269$&$  143$&$ 2.86$&$ 3.69$\\
            $0.6$& $0.80$ &$1.0$&$1.3$&$  266$&$ 2.72$&$ 3.50$&&$0.1408$&$  269$&$ 2.85$&$ 3.81$\\
            $0.6$& $0.95$ &$1.0$&$1.3$&$  611$&$ 2.62$&$ 3.35$&&$0.1534$&$  612$&$ 2.62$&$ 3.43$\\
            $0.6$&$1.00$ &$1.0$&$1.3$&$  864$&$ 2.49$&$ 3.36$&&$0.1574$&$  864$&$ 2.64$&$ 3.49$\\
            $0.6$& $1.05$ &$1.0$&$1.3$&$ 1291$&$ 2.37$&$ 3.17$&&$0.1613$&$ 1292$&$ 2.64$&$ 3.38$\\
            $0.9$& $0.65$ &$1.5$&$1.2$&$  194$&$ 2.58$&$ 3.76$&&$0.1323$&$  204$&$ 2.70$&$ 3.95$\\
            $0.9$& $0.80$ &$1.5$&$1.2$&$  427$&$ 2.53$&$ 3.70$&&$0.1468$&$  434$&$ 2.59$&$ 3.67$\\
            $0.9$& $0.95$ &$1.5$&$1.2$&$ 1255$&$ 2.42$&$ 3.59$&&$0.1599$&$ 1256$&$ 2.53$&$ 3.66$\\
            $0.9$& $1.00$ &$1.5$&$1.2$&$ 2045$&$ 2.36$&$ 3.59$&&$0.1641$&$ 2045$&$ 2.49$&$ 3.55$\\
            $0.9$& $1.05$ &$1.5$&$1.2$&$ 3789$&$ 2.40$&$ 3.35$&&$0.1681$&$ 3793$&$ 2.63$&$ 3.66$\\
            $0.9$& $0.65$ &$1.5$&$1.3$&$  152$&$ 2.78$&$ 3.92$&&$0.1904$&$  160$&$ 2.85$&$ 4.14$\\
            $0.9$& $0.80$ &$1.5$&$1.3$&$  298$&$ 2.61$&$ 3.69$&&$0.2112$&$  303$&$ 2.74$&$ 4.08$\\
            $0.9$& $0.95$ &$1.5$&$1.3$&$  696$&$ 2.65$&$ 3.79$&&$0.2301$&$  697$&$ 2.97$&$ 3.93$\\
            $0.9$& $1.00$ &$1.5$&$1.3$&$  988$&$ 2.48$&$ 3.57$&&$0.2361$&$  988$&$ 2.69$&$ 3.91$\\
            $0.9$& $1.05$ &$1.5$&$1.3$&$ 1481$&$ 2.57$&$ 3.57$&&$0.2420$&$ 1483$&$ 2.61$&$ 3.74$\\
\hline
 \end{tabular} \caption{Empirical estimate of the one-sided type I error at the nominal level of $2.5\%$  based on $10000$ simulated NI trials for design 2: a) $\tau_a=\tau_c=2$,  and
the loss-to-follow-up is exponentially distributed with mean $5$ years ($\delta=0.2$); 
b) $\exp(\beta)$ is used only in the sample size calculation; c) the data are simulated under the assumption that $\lambda_1=\lambda_0M_{r0}$ for the test based on the rate ratio metric, and $\lambda_1=\lambda_0+M_{d0}$ for the test based on the rate difference metric.
 }\label{type1_sim2}
\end{center}
\end{table}

  \begin{table}[h]
\begin{center}
\begin{tabular}{cccc rrrr clc rrr  rcc} \\\hline  
                            &&& \multicolumn{6}{c}{rate ratio based test}   &&    \multicolumn{7}{c}{rate difference based test}  \\\cline{4-9} \cline{11-17}
                         & & & & \multicolumn{4}{c}{total sample size }  &   SIM ($\%$)  &&&  \multicolumn{3}{c}{total sample size}  &&  \multicolumn{2}{c}{SIM ($\%$) at}  \\\cline{5-8}\cline{12-14}\cline{16-17}
$\lambda_0$ & $\exp(\beta)$ & $\kappa$ & $M_{r0}$ & $n_{zr}$ & $n_{rl}$ & $n_r$ & $n_{ru}$ &  at $n_r$  && $M_{d0}$ & $n_{dl}$ & $n_{d}$ & $n_{du}$ && $n_{d}$ & $n_r$ \\\hline 	   
$0.6$&$0.65$&$1.0$&$1.2$&$  182$&$  186$&$  192$&$  194$&$80.47$&&$  0.0882$&$  191$&$  198$&$  200$&&$81.45$&$80.29$\\
$0.6$&$0.80$&$1.0$&$1.2$&$  396$&$  397$&$  412$&$  416$&$80.50$&&$  0.0978$&$  401$&$  416$&$  420$&&$80.29$&$80.65$\\
$0.6$&$0.95$&$1.0$&$1.2$&$ 1143$&$ 1142$&$ 1185$&$ 1197$&$79.25$&&$  0.1066$&$ 1143$&$ 1186$&$ 1198$&&$80.73$&$79.53$\\
$0.6$&$1.00$&$1.0$&$1.2$&$ 1853$&$ 1851$&$ 1921$&$ 1941$&$79.30$&&$  0.1094$&$ 1851$&$ 1921$&$ 1941$&&$79.38$&$79.38$\\
$0.6$&$1.05$&$1.0$&$1.2$&$ 3415$&$ 3410$&$ 3540$&$ 3578$&$79.77$&&$  0.1121$&$ 3412$&$ 3543$&$ 3580$&&$79.82$&$79.61$\\
$0.6$&$0.65$&$1.0$&$1.3$&$  143$&$  145$&$  150$&$  152$&$80.66$&&$  0.1269$&$  150$&$  155$&$  157$&&$81.68$&$80.78$\\
$0.6$&$0.80$&$1.0$&$1.3$&$  276$&$  277$&$  288$&$  290$&$80.55$&&$  0.1408$&$  280$&$  291$&$  293$&&$81.22$&$80.83$\\
$0.6$&$0.95$&$1.0$&$1.3$&$  635$&$  634$&$  658$&$  664$&$80.49$&&$  0.1534$&$  634$&$  658$&$  665$&&$80.73$&$80.73$\\
$0.6$&$1.00$&$1.0$&$1.3$&$  897$&$  894$&$  928$&$  938$&$79.65$&&$  0.1574$&$  894$&$  928$&$  938$&&$79.74$&$79.74$\\
$0.6$&$1.05$&$1.0$&$1.3$&$ 1337$&$ 1333$&$ 1384$&$ 1399$&$80.55$&&$  0.1613$&$ 1334$&$ 1385$&$ 1400$&&$79.94$&$80.64$\\
$0.9$&$0.65$&$1.5$&$1.2$&$  191$&$  194$&$  202$&$  206$&$81.41$&&$  0.1323$&$  203$&$  212$&$  216$&&$82.18$&$81.23$\\
$0.9$&$0.80$&$1.5$&$1.2$&$  423$&$  424$&$  442$&$  452$&$80.03$&&$  0.1468$&$  430$&$  449$&$  458$&&$81.56$&$80.09$\\
$0.9$&$0.95$&$1.5$&$1.2$&$ 1241$&$ 1241$&$ 1294$&$ 1323$&$79.71$&&$  0.1599$&$ 1242$&$ 1295$&$ 1325$&&$80.07$&$79.77$\\
$0.9$&$1.00$&$1.5$&$1.2$&$ 2022$&$ 2021$&$ 2107$&$ 2156$&$79.83$&&$  0.1641$&$ 2021$&$ 2107$&$ 2156$&&$79.81$&$79.81$\\
$0.9$&$1.05$&$1.5$&$1.2$&$ 3743$&$ 3740$&$ 3900$&$ 3993$&$80.24$&&$  0.1681$&$ 3744$&$ 3904$&$ 3997$&&$79.62$&$80.12$\\
$0.9$&$0.65$&$1.5$&$1.3$&$  149$&$  152$&$  158$&$  161$&$80.84$&&$  0.1904$&$  159$&$  166$&$  169$&&$81.82$&$80.66$\\
$0.9$&$0.80$&$1.5$&$1.3$&$  295$&$  296$&$  309$&$  315$&$80.78$&&$  0.2112$&$  301$&$  313$&$  320$&&$81.25$&$80.56$\\
$0.9$&$0.95$&$1.5$&$1.3$&$  689$&$  689$&$  718$&$  734$&$80.37$&&$  0.2301$&$  689$&$  719$&$  735$&&$79.94$&$80.49$\\
$0.9$&$1.00$&$1.5$&$1.3$&$  977$&$  976$&$ 1018$&$ 1042$&$79.80$&&$  0.2361$&$  976$&$ 1018$&$ 1042$&&$79.95$&$79.95$\\
$0.9$&$1.05$&$1.5$&$1.3$&$ 1464$&$ 1462$&$ 1525$&$ 1561$&$80.15$&&$  0.2420$&$ 1464$&$ 1526$&$ 1563$&&$79.90$&$80.45$\\
\hline
 \end{tabular} \caption{Estimated sample size at the nominal $80\%$ power  and simulated power (SIM)  at the specified sample size based on $10000$ NI trials for design $1$: a) $\tau_c=2$,  and the overall dropout rate is $25\%$ ($\delta=0.1438$);
b)  In the data simulation, $\lambda_1=\lambda_0\exp(\beta)$ for both tests based on relative and absolute 
rate difference metrics. 
 }\label{power_sim1}
\end{center}
\end{table}

  \begin{table}[h]
\begin{center}
\begin{tabular}{cccc rrrr ccc rrrcccccccccccc} \\\hline  
                            &&& \multicolumn{6}{c}{rate ratio based test}   &&    \multicolumn{7}{c}{rate difference based test}  \\\cline{4-9} \cline{11-17}
                         & & & & \multicolumn{4}{c}{total sample size}  &   SIM ($\%$)  &&&  \multicolumn{3}{c}{total sample size}  &&  \multicolumn{2}{l}{SIM ($\%$) at}  \\\cline{5-8}\cline{12-14}\cline{16-17}
$\lambda_0$ & $\exp(\beta)$ & $\kappa$ & $M_{r0}$ & $n_{zr}$ & $n_{rl}$ & $n_r$ & $n_{ru}$ &  at $n_r$  && $M_{d0}$ & $n_{dl}$ & $n_{d}$ & $n_{du}$ && $n_{d}$ & $n_r$ \\\hline 	   
$0.6$&$0.65$&$1.0$&$1.2$&$  160$&$  163$&$  176$&$  182$&$80.47$&&$  0.0882$&$  169$&$  183$&$  190$&&$81.86$&$80.18$\\
$0.6$&$0.80$&$1.0$&$1.2$&$  350$&$  351$&$  381$&$  396$&$79.66$&&$  0.0978$&$  355$&$  385$&$  401$&&$80.64$&$79.54$\\
$0.6$&$0.95$&$1.0$&$1.2$&$ 1016$&$ 1016$&$ 1102$&$ 1149$&$80.81$&&$  0.1066$&$ 1016$&$ 1103$&$ 1150$&&$80.06$&$80.80$\\
$0.6$&$1.00$&$1.0$&$1.2$&$ 1650$&$ 1648$&$ 1789$&$ 1868$&$79.97$&&$  0.1094$&$ 1648$&$ 1789$&$ 1868$&&$79.74$&$79.74$\\
$0.6$&$1.05$&$1.0$&$1.2$&$ 3045$&$ 3042$&$ 3302$&$ 3450$&$79.78$&&$  0.1121$&$ 3044$&$ 3304$&$ 3453$&&$80.21$&$79.44$\\
$0.6$&$0.65$&$1.0$&$1.3$&$  125$&$  128$&$  138$&$  143$&$81.15$&&$  0.1269$&$  133$&$  143$&$  149$&&$81.98$&$81.02$\\
$0.6$&$0.80$&$1.0$&$1.3$&$  244$&$  245$&$  266$&$  276$&$79.93$&&$  0.1408$&$  248$&$  269$&$  280$&&$81.12$&$80.27$\\
$0.6$&$0.95$&$1.0$&$1.3$&$  564$&$  564$&$  611$&$  638$&$79.94$&&$  0.1534$&$  564$&$  612$&$  638$&&$79.85$&$80.04$\\
$0.6$&$1.00$&$1.0$&$1.3$&$  798$&$  796$&$  864$&$  902$&$80.00$&&$  0.1574$&$  796$&$  864$&$  902$&&$79.75$&$79.75$\\
$0.6$&$1.05$&$1.0$&$1.3$&$ 1192$&$ 1189$&$ 1291$&$ 1349$&$81.22$&&$  0.1613$&$ 1190$&$ 1292$&$ 1350$&&$80.13$&$80.94$\\
$0.9$&$0.65$&$1.5$&$1.2$&$  176$&$  178$&$  194$&$  208$&$79.88$&&$  0.1323$&$  188$&$  204$&$  220$&&$82.48$&$79.66$\\
$0.9$&$0.80$&$1.5$&$1.2$&$  392$&$  394$&$  427$&$  460$&$80.35$&&$  0.1468$&$  400$&$  434$&$  468$&&$80.64$&$80.30$\\
$0.9$&$0.95$&$1.5$&$1.2$&$ 1157$&$ 1157$&$ 1255$&$ 1357$&$79.86$&&$  0.1599$&$ 1158$&$ 1256$&$ 1358$&&$80.38$&$80.18$\\
$0.9$&$1.00$&$1.5$&$1.2$&$ 1887$&$ 1886$&$ 2045$&$ 2215$&$80.10$&&$  0.1641$&$ 1886$&$ 2045$&$ 2215$&&$80.01$&$80.01$\\
$0.9$&$1.05$&$1.5$&$1.2$&$ 3497$&$ 3495$&$ 3789$&$ 4108$&$80.86$&&$  0.1681$&$ 3499$&$ 3793$&$ 4112$&&$80.33$&$80.74$\\
$0.9$&$0.65$&$1.5$&$1.3$&$  138$&$  140$&$  152$&$  162$&$80.23$&&$  0.1904$&$  148$&$  160$&$  172$&&$82.24$&$80.12$\\
$0.9$&$0.80$&$1.5$&$1.3$&$  274$&$  275$&$  298$&$  321$&$80.23$&&$  0.2112$&$  279$&$  303$&$  327$&&$80.77$&$80.19$\\
$0.9$&$0.95$&$1.5$&$1.3$&$  642$&$  642$&$  696$&$  753$&$80.05$&&$  0.2301$&$  642$&$  697$&$  754$&&$80.02$&$80.09$\\
$0.9$&$1.00$&$1.5$&$1.3$&$  912$&$  911$&$  988$&$ 1070$&$80.23$&&$  0.2361$&$  911$&$  988$&$ 1070$&&$80.18$&$80.18$\\
$0.9$&$1.05$&$1.5$&$1.3$&$ 1368$&$ 1367$&$ 1481$&$ 1606$&$80.13$&&$  0.2420$&$ 1368$&$ 1483$&$ 1608$&&$79.63$&$79.94$\\
\hline
 \end{tabular} \caption{Estimated sample size at the nominal $80\%$ power  and simulated power (SIM) at the specified sample size based on $10000$ NI trials for design $2$: a) $\tau_a=\tau_c=2$,  and
the loss-to-follow-up is exponentially distributed with mean $5$ years ($\delta=0.2$); b)  In the data simulation, $\lambda_1=\lambda_0\exp(\beta)$ for both tests based on relative and absolute 
rate difference metrics. 
 }\label{power_sim2}
\end{center}
\end{table}

  \begin{table}[h]
\begin{center}
\begin{tabular}{ccccc rrrc ccrrrrcccccccccccc} \\\hline  
                       &     &&& \multicolumn{5}{c}{rate ratio based test}   &&    \multicolumn{7}{c}{rate difference based test}  \\\cline{5-9} \cline{11-17}
                        & & & & & \multicolumn{3}{c}{total sample size}  &   SIM ($\%$)  &&&  \multicolumn{3}{c}{total sample size}  &&  \multicolumn{2}{l}{SIM ($\%$) at}  \\\cline{6-8}\cline{12-14}\cline{16-17}
$\lambda_0$ & $\kappa_0$ & $\lambda_1$ & $\kappa_1$ & $M_{r0}$  & $n_{rl}$ & $n_r$ & $n_{ru}$ &  at $n_r$  && $M_{d0}$ & $n_{dl}$ & $n_{d}$ & $n_{du}$ && $n_{r}$ & $n_d$ \\\hline 
$0.6$&$ 2.0$&$0.48$&$ 1.0$&$1.3$&$  344$&$  358$&$  363$&$79.48$&&$  0.1408$&$  363$&$  378$&$  384$&&$79.66$&$82.19$\\
$0.6$&$ 1.0$&$0.48$&$ 2.0$&$1.3$&$  344$&$  358$&$  363$&$80.95$&&$  0.1408$&$  333$&$  347$&$  351$&&$80.83$&$79.70$\\
$0.6$&$ 2.0$&$0.48$&$ 0.5$&$1.3$&$  311$&$  322$&$  327$&$79.34$&&$  0.1408$&$  337$&$  349$&$  355$&&$79.38$&$82.64$\\
$0.6$&$ 0.5$&$0.48$&$ 2.0$&$1.3$&$  311$&$  322$&$  327$&$80.62$&&$  0.1408$&$  292$&$  302$&$  306$&&$80.44$&$78.67$\\
$1.0$&$ 2.0$&$0.80$&$ 1.0$&$1.3$&$  286$&$  298$&$  306$&$79.71$&&$  0.2347$&$  306$&$  319$&$  327$&&$79.85$&$82.46$\\
$1.0$&$ 1.0$&$0.80$&$ 2.0$&$1.3$&$  286$&$  299$&$  306$&$80.62$&&$  0.2347$&$  276$&$  288$&$  294$&&$80.72$&$79.00$\\
$1.0$&$ 2.0$&$0.80$&$ 0.5$&$1.3$&$  253$&$  263$&$  269$&$78.69$&&$  0.2347$&$  279$&$  290$&$  298$&&$79.18$&$83.69$\\
$1.0$&$ 0.5$&$0.80$&$ 2.0$&$1.3$&$  253$&$  263$&$  269$&$81.22$&&$  0.2347$&$  234$&$  244$&$  249$&&$81.03$&$78.19$\\
$0.6$&$ 2.0$&$0.54$&$ 1.0$&$1.3$&$  584$&$  607$&$  617$&$79.36$&&$  0.1493$&$  598$&$  622$&$  632$&&$80.16$&$81.28$\\
$0.6$&$ 1.0$&$0.54$&$ 2.0$&$1.3$&$  584$&$  608$&$  617$&$80.31$&&$  0.1493$&$  573$&$  597$&$  606$&&$80.13$&$79.29$\\
$0.6$&$ 2.0$&$0.54$&$ 0.5$&$1.3$&$  526$&$  545$&$  553$&$78.85$&&$  0.1493$&$  546$&$  566$&$  575$&&$80.31$&$81.64$\\
$0.6$&$ 0.5$&$0.54$&$ 2.0$&$1.3$&$  526$&$  546$&$  553$&$80.78$&&$  0.1493$&$  509$&$  528$&$  535$&&$80.02$&$78.71$\\
$1.0$&$ 2.0$&$0.90$&$ 1.0$&$1.3$&$  490$&$  510$&$  523$&$79.55$&&$  0.2489$&$  504$&$  525$&$  538$&&$80.27$&$82.04$\\
$1.0$&$ 1.0$&$0.90$&$ 2.0$&$1.3$&$  490$&$  510$&$  523$&$80.83$&&$  0.2489$&$  479$&$  499$&$  512$&&$80.35$&$79.26$\\
$1.0$&$ 2.0$&$0.90$&$ 0.5$&$1.3$&$  432$&$  449$&$  459$&$78.74$&&$  0.2489$&$  452$&$  469$&$  481$&&$80.12$&$82.79$\\
$1.0$&$ 0.5$&$0.90$&$ 2.0$&$1.3$&$  432$&$  449$&$  459$&$81.04$&&$  0.2489$&$  415$&$  431$&$  441$&&$80.04$&$78.66$\\
$0.6$&$ 2.0$&$0.60$&$ 1.0$&$1.3$&$ 1122$&$ 1168$&$ 1187$&$80.13$&&$  0.1574$&$ 1122$&$ 1168$&$ 1187$&&$81.40$&$81.40$\\
$0.6$&$ 1.0$&$0.60$&$ 2.0$&$1.3$&$ 1122$&$ 1168$&$ 1187$&$80.16$&&$  0.1574$&$ 1122$&$ 1168$&$ 1187$&&$79.13$&$79.13$\\
$0.6$&$ 2.0$&$0.60$&$ 0.5$&$1.3$&$ 1008$&$ 1046$&$ 1063$&$78.82$&&$  0.1574$&$ 1008$&$ 1046$&$ 1063$&&$80.82$&$80.82$\\
$0.6$&$ 0.5$&$0.60$&$ 2.0$&$1.3$&$ 1008$&$ 1046$&$ 1063$&$80.41$&&$  0.1574$&$ 1008$&$ 1046$&$ 1063$&&$78.60$&$78.60$\\
$1.0$&$ 2.0$&$1.00$&$ 1.0$&$1.3$&$  947$&$  987$&$ 1012$&$80.23$&&$  0.2624$&$  947$&$  987$&$ 1012$&&$81.20$&$81.20$\\
$1.0$&$ 1.0$&$1.00$&$ 2.0$&$1.3$&$  947$&$  987$&$ 1012$&$80.24$&&$  0.2624$&$  947$&$  987$&$ 1012$&&$78.95$&$78.95$\\
$1.0$&$ 2.0$&$1.00$&$ 0.5$&$1.3$&$  833$&$  866$&$  888$&$79.06$&&$  0.2624$&$  833$&$  866$&$  888$&&$81.67$&$81.67$\\
$1.0$&$ 0.5$&$1.00$&$ 2.0$&$1.3$&$  833$&$  866$&$  888$&$80.92$&&$  0.2624$&$  833$&$  866$&$  888$&&$78.82$&$78.82$\\
\hline
 \end{tabular} \caption{Estimated sample size at the nominal $80\%$ power  and simulated power (SIM)  at the specified sample size based on $10000$ NI trials for design $1$ when the dispersion parameter varies by treatment groups: a)  $\tau_c=2$,  and the overall dropout rate is $25\%$ ($\delta=0.1438$).
 }\label{power_herdisp}
\end{center}
\end{table}

  \begin{table}[h]
\begin{center}
\begin{tabular}{cccc rrrc clc rcc  ccc} \\\hline  
                            &&& \multicolumn{5}{c}{rate ratio based test}   &&    \multicolumn{4}{c}{rate difference based test}  \\\cline{4-8} \cline{10-13}
                         & & &  \multicolumn{4}{c}{total sample size }  &   SIM ($\%$)  &&  \multicolumn{3}{c}{total sample size}  &  SIM ($\%$)   \\\cline{4-7}\cline{10-12}
$\lambda_0$ & $\exp(\beta)$ & $\kappa$  & $n_{zr}$$^{(a)}$ & $n_{rl}$ & $n_r$ & $n_{ru}$ &  at $n_r$  &&  $n_{dl}$ & $n_{d}$ & $n_{du}$ & at $n_{d}$ \\\hline 	   
\multicolumn{12}{l}{Design I: $\tau_c=2$, overall dropout rate $25\%$ ($\delta=0.1438$)}\\
$0.6$&$1.00$&$1.0$&$  1200$&$  1197$&$  1242$&$  1255$&$79.83$&& $  1197$&$  1242$&$  1255$&$80.12$\\
$0.6$&$1.05$&$1.0$&$ 1386$&$ 1382$&$ 1435$&$ 1451$&$79.92$&& $ 1383$&$ 1436$&$ 1452$&$80.07$\\
$0.9$&$1.00$&$1.5$&$ 1308$&$  1307$&$ 1363$&$ 1394$&$79.75$&&$  1307$&$ 1363$&$ 1394$&$79.74$\\
$0.9$&$1.05$&$1.5$&$ 1518$&$ 1516$&$ 1581$&$ 1619$&$80.17$&& $ 1518$&$ 1583$&$ 1620$&$80.10$\\
\vspace{0.02in}\\
\multicolumn{12}{l}{Design II: $\tau_a=\tau_c=2$, exponential dropout rate $\delta=0.2$}\\
$0.6$&$1.00$&$1.0$&$ 1068$&$  1066$&$  1157$&$  1208$&$80.28$&& $  1066$&$  1157$&$  1208$&$80.17$\\
$0.6$&$1.05$&$1.0$&$ 1236$&$ 1233$&$ 1339$&$ 1399$&$79.75$&& $ 1234$&$ 1340$&$ 1400$&$80.41$\\
$0.9$&$1.00$&$1.5$&$ 1190$&$  1189$&$ 1288$&$ 1402$&$79.50$&&$  1189$&$ 1288$&$ 1402$&$79.28$\\
$0.9$&$1.05$&$1.5$& $ 1418$&$ 1417$&$ 1536$&$ 1666$&$79.94$&&$ 1418$&$ 1538$&$ 1667$&$80.03$\\
\hline
 \end{tabular} \caption{Estimated sample size at the nominal $80\%$ power  and simulated power (SIM)  at the specified sample size based on $10000$ equivalence trials: 
a) $n_{zr}$ is estimated using method $3$ in Section $4$ of \cite{zhu:2016};
b)  In the data simulation, $\lambda_1=\lambda_0\exp(\beta)$ for both tests based on relative and absolute rate difference metrics;
c) The margin is $M_{ru}=1/M_{rl}=1.3$ on the rate ratio metric and $M_{du}=-M_{dl}=\exp(\beta/2)\lambda_0 \log(M_{ru})  $ on the absolute rate difference metric;
d) the lower and upper size bounds are obtained by numerically inverting the power bounds.
 }\label{power_eqivsim}
\end{center}
\end{table}

\end{document}